\documentclass[twocolumn,dvipsname]{aastex631}

\let\tablenum\relax
\usepackage{siunitx}
\usepackage{booktabs}
\usepackage{makecell}
\usepackage{xspace}
\usepackage{amsmath}
\usepackage{multirow}
\usepackage{float}
\let\oldAA\AA
\renewcommand{\AA}{{\text{\normalfont\oldAA}}\xspace}

\graphicspath{{./}{figures/}}

\def\apjs{{ApJS}}

\def\h{\hskip -3 mm}

\def\lax{{$\mathrel{\hbox{\rlap{\hbox{\lower4pt\hbox{${\sim}$}}}\hbox{$<$}}}$}}
\def\gax{{$\mathrel{\hbox{\rlap{\hbox{\lower4pt\hbox{${\sim}$}}}\hbox{$>$}}}$}}
\def\simlt{\lower.5ex\hbox{$\; \buildrel < \over {\sim} \;$}}
\def\simgt{\lower.5ex\hbox{$\; \buildrel > \over {\sim} \;$}}

\def\logm10{{$\log (M_{\star,10\ \mathrm{kpc}}/M_{\odot})$}}
\def\logm30{{$\log (M_{\star,30\ \mathrm{kpc}}/M_{\odot})$}}
\def\logm50{{$\log (M_{\star,50\ \mathrm{kpc}}/M_{\odot})$}}
\def\logm100{{$\log (M_{\star,100\ \mathrm{kpc}}/M_{\odot})$}}

\def\mh200b{{$M_{\mathrm{200b}}$}}
\def\mh200c{{$M_{\mathrm{200c}}$}}

\def\mdpl2{\texttt{MDPL2}}

\definecolor{LightGray}{gray}{0.85}
\definecolor{Tab1}{RGB}{114, 158, 206}
\definecolor{Tab2}{RGB}{255, 158,  74}
\definecolor{Tab3}{RGB}{103, 191,  92}
\definecolor{Tab4}{RGB}{174, 199, 232}
\definecolor{Tab5}{RGB}{255, 187, 120}
\definecolor{Tab6}{RGB}{152, 223, 138}
\definecolor{Tab7}{RGB}{255, 152, 150}
\definecolor{Tab8}{RGB}{197, 176, 213}
\definecolor{hpurple}{HTML}{7E16DF}

\begin{document}

\title{Oxyster: A Circumgalactic Low-ionization Oxygen Nebula next to a Starburst Galaxy at $z\sim1$}

\author[0009-0000-4701-4934]{Pengjun Lu}
\affiliation{Department of Astronomy, Tsinghua University, Beijing 100084, China}

\author[0000-0001-6251-649X]{Mingyu Li}
\affiliation{Department of Astronomy, Tsinghua University, Beijing 100084, China}

\author[0000-0003-4974-3481]{Dalya Baron}
\affiliation{Kavli Institute for Particle Astrophysics \& Cosmology (KIPAC), Stanford University, CA 94305, USA}
\affiliation{Center for Decoding the Universe, Stanford University, CA 94305, USA}

\author[0000-0002-5367-8021]{Minghao Yue}
\affiliation{MIT Kavli Institute for Astrophysics and Space Research, MA 02139, USA}

\author[0000-0003-1385-7591]{Song Huang}
\affiliation{Department of Astronomy, Tsinghua University, Beijing 100084, China}

\author[0000-0001-8467-6478]{Zheng Cai}
\affiliation{Department of Astronomy, Tsinghua University, Beijing 100084, China}


\correspondingauthor{\href{mailto:lupj23@mails.tsinghua.edu.cn, lmytime@hotmail.com, shuang@mail.tsinghua.edu.cn}{Pengjun Lu, Mingyu Li, Song Huang}}
\email{lupj23@mails.tsinghua.edu.cn}

\begin{abstract}
    Extended emission line nebulae around galaxies or active galactic nuclei (AGNs) provide a unique window to investigate the galactic ecosystem through the circumgalactic medium (CGM). Using Subaru Hyper-Suprime Cam narrow-band imaging and spectroscopic follow-up, we serendipitously discover ``Oxyster" --- a large ionized nebula next to an interacting starburst galaxy at $z=0.924$. The nebula is traced by extended [\ion{O}{2}]$3726,3729\AA$ ($\sim 30$ kpc) and [\ion{O}{3}]$5007\AA$ ($\sim 20$ kpc) emission lines. On the nebula luminosity-size plane, Oxyster surpasses the extended narrow-line regions around low-$z$ AGNs, resembling a higher-$z$ analog of ``Hanny's Voorwerp". However, its uniformly low [\ion{O}{3}]/[\ion{O}{2}] ratio (O32) sets it apart from typical AGN light echoes. For the host galaxy, HST and JWST images reveal a disturbed red disk galaxy with a single blue spiral ``arm". Spectral energy distribution (SED) fitting suggests the $2-6\times 10^{10} ~\rm M_{\odot}$ host galaxy sits above the star-forming main sequence with an ongoing starburst, especially in the ``arm", and have $<5$\% luminosity contribution from AGN, consistent with X-ray non-detection and radio continuum. Standard photoionization and shock models struggle to explain simultaneously Oxyster's emission line luminosities, low O32 ratio, and the non-detection of H$\beta$ line. A plausible explanation could involve the combination of a recent ($<10^8$ yrs) starburst and a low-luminosity AGN ($L_{\rm{bol}} \sim 1\times10^{42}$ erg/s). While Oxyster's nature awaits future investigation, its discovery highlights the potential of ground-based narrow-band imaging to uncover extended emission line nebulae around non-AGN systems, opening new avenues for studying the CGM of normal galaxies in the early Universe.

\end{abstract}



\section{Introduction} 
    \label{sec:intro}

    In recent years, one of the long-standing challenges in astronomy is to understand the baryonic cycle in the galactic ecosystem \citep[e.g.,][]{Peroux-2020, Donahue-2022}, which describes a broad range of intricately connected physical processes, such as the gas accretion, star formation, supernova feedback, large-scale outflow driven by a supermassive black hole (SMBH), all of which are fundamental to galaxy evolution \citep[e.g.,][]{Angles-Alcazar-2017, Faucher-Giguere-2023, Nishigaki-2025}. Among them, the stellar and active galactic nucleus (AGN) feedback are of particular interest now, as they can regulate the gas accretion \& star formation (or the quenching of it) in galaxies and can provide the crucial channels to transport energy, angular momentum, and metals formed in stars across the interstellar medium (ISM) or even the circumgalactic medium (CGM) through ionizing radiation and powerful winds \citep[e.g.,][]{King-2015, Harrison-2018, Zhang-2018}. 

    Among the many direct and indirect observational evidence of such critical processes, extended emission line regions around different types of galaxies, either as a ``halo'' centered on the galaxy or as a separated ``nebula'',  present the rare and valuable opportunities to actually \emph{see} the cold-phase of the CGM.
    The gas has been photo- or collisional-ionized by some recent events in the last 10-100 Myrs, allowing us to visualize the baryonic cycle temporarily \citep[e.g.][]{Tumlinson-2011, Lokhorst-2022, Lin-2023, Zhang-2023}. Thanks to the advances of deep imaging and integrated-field spectroscopic/unit (IFS/IFU) observations, we can directly measure the physical extent, velocity structure, gas metallicity, and strength of the ionizing radiation field, among other physical characteristics of the CGM. However, given the diffuse nature of the CGM and the rarity of an intense ionizing event, such emission line halos or nebulae are often short-lived and challenging to observe. 
    The strategy of stacking a large sample of objects helps reveal the underlying average properties of the CGM in emission lines \citep[e.g.,][]{Comparat-2022, Jones-2023, Dutta-2024, Romano-2024}. Still, it falls short of depicting the crucial details of these critical physical processes. As for the detailed study of individual cases, high-energy ionizing sources such as high-luminosity quasars or AGNs are often favored \citep[e.g.,][]{Cantalupo2008ApJ...672...48C, ArrigoniBattaia2018MNRAS.473.3907A, Johnson-2018, DeBreuck-2022, Venturi-2023, Johnson-2024, Vayner-2024, Liu-2025}, as these extraordinary objects can shine through the CGM to $>100$ kpc \citep[e.g.,][]{ArrigoniBattaia2019MNRAS.482.3162A, Peng-2025}. 

    While bright quasars or AGNs are spectacular, the emission line regions around less active or more ``regular'' galaxies, such as low-luminosity AGN, dusty star-forming galaxies, post-starburst systems, or even quiescent galaxies \citep[e.g.,][]{Prieto-2016, French-2023, Wevers-2024, Solimano2025A&A...693A..70S, D'Eugenio2025arXiv250315590D}, are equally important, as they are often particularly revealing about a specific critical physical process in galaxy evolution. For instance, the famous ``Hanny's Voorwerp'' is a giant [\ion{O}{3}]$\lambda$4959,5007$\AA$ emission line nebula extended beyond 20 kpc away from a local spiral galaxy IC~2497 at $z=0.05$ \citep[e.g.,][]{Lintott-2009}. The physical properties of ``Hanny's Voorwerp'' unambiguously illustrate a case where the central AGN of IC~2497 has been effectively ``switched off'' in the past $10^5$ yr \citep[e.g.,][]{Schawinski-2010, Keel-2012, Sartori-2016}. Although as indirect evidence, it provides one of the best observational constraints on AGNs' duty cycle at low redshift and inspires the search \& discoveries of similar cases of ``AGN light echoes'' (e.g., \citealt{Schweizer-2013, Keel-2015}). Similarly, using the extended nebula emission around a post-starburst galaxy, \citet{Baron-2018} revealed the direct evidence that AGN-driven winds can effectively strip gas away from the galaxy, capturing an essential moment in the quenching process and supporting the crucial role played by the central SMBH in it. 

    Until now, most of these emission line nebula around ``regular'' galaxies are found in the local Universe, while AGN and star forming activities are wilder toward higher redshift \citep[e.g.][]{Ueda-2003, Aird-2015, Fotopoulou-2016, Kulkarni-2019}, there should be more shining remnants around high-redshift galaxies await to be discovered. Such cases will open a new window to probe the kinematic composition, ionizing source, and physical connection of the ISM \& CGM in galaxies at earlier epochs when gas accretion was higher, starburst was more intense, and AGN activities were more common. 

    Luckily, with the advance of deep narrow/medium-band ground-based imaging surveys, which often target high-redshift Ly$\alpha$ emitters (LAEs) and Ly$\alpha$ nebula around bright quasars \citep[e.g.,][]{Steidel2000ApJ...532..170S, Yang2009ApJ...693.1579Y, Cai2017ApJ...837...71C, Zhang2020ApJ...891..177Z, Li2024ApJS..275...27L}, extended nebula traced by [\ion{O}{2}]$\lambda$3727,3729$\AA$, H$\beta$$\lambda$4861$\AA$, or [\ion{O}{3}]$\lambda$4959,5007$\AA$ emission lines can already be discovered to $z\sim1$ \citep[e.g.,][]{Yuma2013ApJ...779...53Y, Yuma2017ApJ...841...93Y, Sun-2018}.  

    Here, we report the serendipitous discovery of a compelling emission line nebula system surrounding a star-forming galaxy at \(z = 0.924\), identified through the Hyper Suprime-Cam (HSC; \citealt{Miyazaki-2018}) narrow-band images in the Cosmic Evolution Survey \citep[COSMOS; e.g.,][]{Scoville-2007} field. Both the NB718 and NB973 images reveal a giant nebula extending up to 30 kpc, revealing emissions from both the [\ion{O}{2}] and [\ion{O}{3}] lines. Neither the spectral energy distribution (SED) fitting nor the multiwavelength data reveals evidence of an ongoing AGN in the host galaxy. And, unlike the ``AGN's light echoes'' at low-redshift, this system has a low [\ion{O}{3}]$\lambda 5007\AA$/[\ion{O}{2}]$\lambda 3726,3729\AA$ ratio ($<1$), suggesting a low-ionization state. The host galaxy is experiencing a moderate starburst and displays an asymmetric spiral structure, possibly induced by the interaction of a nearby system. Our discovery offers a vivid example of how such an extended emission line nebula can provide valuable insights into the interaction between the galaxies and their CGMs at $z\sim1$.

    In Section \ref{sec:data}, we describe the data used. Section \ref{sec:method} outlines the photometric and spectroscopic processing methods, leading to the summary of our results in Section \ref{sec:results}. Section \ref{sec:discuss} discusses the potential ionization sources illuminating the nebula and the implications of the host galaxy's properties. Finally, in Section \ref{sec:conclude}, we summarize our findings and suggest directions for future observations to test our hypotheses further. Throughout this paper, we adopt a flat $\Lambda$CDM cosmological model with $H_0 = 70~\mathrm{km~s^{-1}~Mpc^{-1}}$, $\Omega_M = 0.3$, and $\Omega_\Lambda = 0.7$.

\section{The Discovery and Multiwavelength Data}
    \label{sec:data}

    In this section, we present the serendipitous discovery and multiwavelength investigation of a circumgalactic [\ion{O}{2}] nebula extending out to $\sim30$~kpc at $z=0.924$ in the COSMOS field, which is associated with a massive star-forming galaxy that shows no clear evidence of AGN.\footnote{\textbf{We notice that Oxyster was individually identified by A. F. Jonkeren and reported in the \href{https://www.zooniverse.org/projects/zookeeper/galaxy-zoo/talk/1269/858076?page=108}{GalaxyZoo's ``Voorwerpje \& Green Bean Hunt" discussion forum} along with many other interesting candidates. We acknowledge the contribution from the GalaxyZoo community.}}
    We initially came across this system when visually inspecting Subaru HSC narrow-band images \citep{Aihara2022PASJ...74..247A, Inoue-2020} in the COSMOS field to identify enormous Lyman-$\alpha$ nebulae at $z>2$ (Li et al. in prep.).
    A more detailed analysis and further spectroscopic follow-up confirm that the extreme flux excess in the NB718 and NB973 narrow-band images (see Figure \ref{RGBfigs}) should come from the [\ion{O}{2}] and [\ion{O}{3}] emission lines at $z=0.924$ (see Section \ref{sec:spectra}). This nebula looks like a giant oyster shining in Oxygen line emission, so we name this system \emph{Oxyster}.
    
    Thanks to the plethora of multiwavelength data accumulated over the years in the COSMOS field, we can provide a detailed and quantitative description of the properties of \emph{Oxyster}, especially with imaging observation by the James Webb Space Telescope \citep[JWST;][]{Gardner-2006}. We carried out a thorough spectral energy distribution (SED) analysis to infer the physical properties of the galaxy and explored archival radio to X-ray data to shed light on the nature of the energy source that powers the nebula. In addition, we took a long-slit, low-resolution optical spectrum of \emph{Oxyster} to confirm its redshift and extract basic information about the emission line gas.
    This section provides an overview of the multiwavelength data used in this work, summarized in Table \ref{photodata} and shown in the appendix \ref{sec:appendix}.

\begin{table*}[htbp]
\centering
\begin{tabular}{ccc}
\hline
Telescope & Instrument/Band                                                                                                      & Survey     \\ \hline
HST       & ACS/F606W$^*$, F814W$^*$, WFC3/F125W, F160W                                                                                                     & CANDELS    \\
CFHT      & Mega-Prime/i, u WIRCam/H, Ks                                                                                         & CFHT Data   \\
Subaru    & \begin{tabular}[c]{@{}c@{}}HSC/g, r, i, z, Y, IB945, NB527, \\ NB718, NB816, NB921, NB973\end{tabular} & CHORUS \& HSC-SSP     \\
JWST      & NIRCam/F115W$^*$, F150W$^*$, F277W$^*$, F444W$^*$                                                                                    & COSMOS-Web \\ \hline
\end{tabular}
\caption{Data used in galaxy SED photometry.}
\tablecomments{$^*$The bands used for measuring the photometry of separate ``arm'' and ``core'' structures.}
\label{photodata}
\end{table*}

\subsection{Imaging Data} 
    \label{Sec:COSMOS}

\subsubsection{Subaru/HSC Images}

    Subaru/HSC provides most of the high-quality rest-frame UV-optical images used in this work, which were taken under $<1.0$ arcsec seeing conditions. Specifically, we utilized the narrow-band images from the PDR1 of CHORUS \citep[][]{Inoue-2020} program that involves the NB387, NB527, NB718, IB945, NB973, and NB921 filters centered at 386.2 nm, 526.0 nm, 717.1 nm, 946.2 nm, and 971.2 nm separately. We also used HSC images at broad bands g, r, i, z, and Y to extract the galaxy continuum and SED. \emph{Oxyster} appears only in the narrow band NB718 and NB973 filters that coincide with [\ion{O}{2}] $\lambda\lambda 3726,3729 \AA$ and [\ion{O}{3}] $\lambda 5007 \AA$ emission at redshift $z\sim0.9$. Meanwhile, the spectroscopic redshift of the galaxy immediately next to \emph{Oxyster} is $z_\mathrm{spec}=0.924$ from the medium-resolution Deep Imaging Multi-Object Spectrograph \citep[DEIMOS;][]{Hasinger-2018}, suggesting that the galaxy could be the physical host of \emph{Oxyster}. 
    
    Therefore, we suspected that the giant nebula and the galaxy are physically connected. We also seek alternative explanations by checking all possible pairs of strong emission lines that fall into these two narrow bands. In principle, a bright $z=4.9$ QSO could have its Ly$\alpha$ and \ion{He}{2} lines meet this criterion \citep{Vanden_Berk_2001}, but then its strong \ion{C}{4} emission should have also appeared in the NB921 filter as well. Moreover, Oxyster appears to be extended and diffuse in both narrow-band images without a clear point source, making the QSO scenario even more unlikely. These lines are also consistent with emission from Population \uppercase\expandafter{\romannumeral3} stars, but the required line luminosities would exceed predictions from models \citep[e.g.][]{Mas-Ribas-2016, Venditti-2024}. 
    
    Furthermore, on the ZFOURGE \citep{Straatman-2016} median J-band image, we find a tentative detection of a diffuse emission corresponding to Oxyster, consistent with being the H$\alpha$ line at $z\sim0.9$, supporting the oxygen emission line nebula interpretation further. 

\subsubsection{The JWST and HST Images}

    While the Subaru/HSC images are sufficiently deep to discover Oxyster, they lack the spatial resolution to resolve the host galaxy's structures. Fortunately, Oxyster's location is within the coverage of the Hubble Space Telescope's ({\it HST}) Cosmic Assembly Near-infrared Deep Extragalactic Legacy Survey (CANDELS, \cite{Grogin-2011, Koekemoer-2011}). CANDELS provides high-resolution images in four filters (F606W, F814W, F125W, and F160W). At $z\sim0.9$, {\it HST}'s \si{{0.05}\arcsecond} per pixel resolution corresponds to $\sim 0.4$ kpc physical scale, capturing the detailed structures of the host galaxy between the rest-frame 315 nm and 423 nm - a wavelength range sensitive to the ongoing star formation activity in the galaxy. 
    
    At the same time, James Webb Space Telescope ({\it JWST}; \cite{Gardner-2006}) COSMOS-Web Survey \citep{Casey-2023} recently released 0.54 deg$^2$ of NIRCam mosaic images in F115W, F150W, F277W, and F444W filters, corresponding to the rest-frame V, I, H, and K-band at $z\sim0.9$. These JWST data provide an excellent opportunity to study the rest-frame optical to NIR properties of Oxyster's host galaxy and constrain its star formation history (SFH) and dust attenuation through multi-band SED fitting. We generate the cutout images of Oxyster using the NIRCam mosaic images and the tools provided by the Dawn JWST Archive (DJA\footnote{\url{https://dawn-cph.github.io/dja/general/mapview/}}; \citealt{Valentino-2023}). These data were reduced using the \texttt{grizli} \citep{Brammer-2023} pipeline to a uniform pixel scale of $0\farcs05$. \textbf{The data described here may be obtained from the MAST archive at
    \dataset[doi:10.17909/T94S3X]{https://dx.doi.org/10.17909/T94S3X} and \dataset[doi:10.17909/fbtm-a977]{https://dx.doi.org/10.17909/fbtm-a977}.}

\subsubsection{Other Photometric Data}

    In addition to the Subaru/HSC-SSP+CHORUS, {\it HST}/CANDELS, and {\it JWST}/COSMOS-Web images, we also supplement the multi-band dataset with the Canada France Hawaii Telescope (CFHT) images tracing the rest-frame UV-to-NIR wavelength range, including the $u$- \& $i$-band data using the Mega-Prime camera and the H \& Ks-band data from the COSMOS-WIRCam Near-Infrared Imaging Survey \citep{McCracken-2010}. 

\subsubsection{Radio \& X-ray Data}

    Moreover, the deep radio \& X-ray data in the COSMOS field could provide valuable insights about the host galaxy's star formation and AGN activities, potentially critical when discussing the physical origin of Oxyster. 

    In the VLA-COSMOS Survey \citep{Schinnerer-2004, Schinnerer-2007, Bondi-2008}, Oxyster's host galaxy is detected in both the 1.4GHz and 3GHz, with the corresponding integrated flux densities of $134~\rm{\mu Jy}$ (from the COSMOS VLA Deep Catalog \citealt{Schinnerer-2010}) and $98.1~\rm{\mu Jy}$ (from the VLA-COSMOS 3-GHz Large Project Source Catalog \citealt{Delvecchio-2017}). On the X-ray side, Oxyster's host galaxy is not detected in the Chandra COSMOS Survey \citep{Civano-2016}, providing the upper limit of soft X-ray flux density of $1.9 \times 10^{-16}~\rm erg~s^{-1}~cm^{-2}$
    
    VLA-COSMOS 3-GHz Large Project Source Catalog classified Oxyster's host galaxy as a ``low-to-moderate radiative luminosity AGN". This classification was based on the radio excess relative to the radio luminosity predicted based on the FIR-radio correlation for star-forming galaxies and a $51~\rm M_{\odot}/yr$ SFR estimate. Later, we will show that the galaxy's SFR is likely to be underestimated by 2 times based on the high-quality SED with {\it HST} $+$ {\it JWST} data. \citet{Delvecchio-2017} also concluded that there is no evidence of AGN activity in Oxyster's host using X-ray \& MIR data or based on the multi-wavelength SED decomposition. We will discuss this further in Section \ref{sec:AGN-ionize}.


\subsection{Spectroscopic Observation}
    \label{sec:spectra}
    
    To confirm the redshift of Oxyster and understand the nature of the nebula emission, we took advantage of an opportunity to observe ``Oxyster'' using the Low-Dispersion Survey Spectrograph (LDSS-3) on the 6.5-m Magellan/Clay telescope as an incidental ``filler'' for a different program when the observational condition was not ideal. We successfully obtained a long-slit spectrum of the nebula with the ``VPH Red" grism that covers the 6070-10760 $\AA$ wavelength range with a $\sim$ 8$\AA$ spectral resolution. We placed a \si{{1.5}\arcsecond} width long slit at a position angle of P.A. = \ang{132} across the galaxy along with the nebula (RA=10:00:38.354, DEC=+02:11:28.198). We accumulate 1-hour useful data from three 20-minute exposure, with ABA nodding with a 25-pixel offset along the slit between the exposures. 

    We reduced the spectra using \texttt{PypeIt} \citep{Prochaska-2020a}, including flat-field correction, wavelength calibration, cosmic ray subtraction, and telluric line subtraction. The data were then co-added in ABA mode, following the flux calibration with the standard star LTT~3864. Given the limited observational condition and the suboptimal instrument setup that was available to us, we only detected an unresolved [\ion{O}{2}]$\lambda\lambda 3726,3729$ doublet and the [\ion{O}{3}]$\lambda 5007$ line. The weaker [\ion{O}{3}]$\lambda 4959$ was not detected. However, the spectrum confirms the redshift of the nebula estimated based on the narrowband data and the physical connection between the nebula and the galaxy.


\section{Photometric Analysis and SED Fitting}
    \label{sec:method}

\subsection{Multi-band Photometry of the Host Galaxy}
    \label{subsec:photometry}

\begin{figure*}[htbp]
    \centering
    \includegraphics[width=\textwidth]{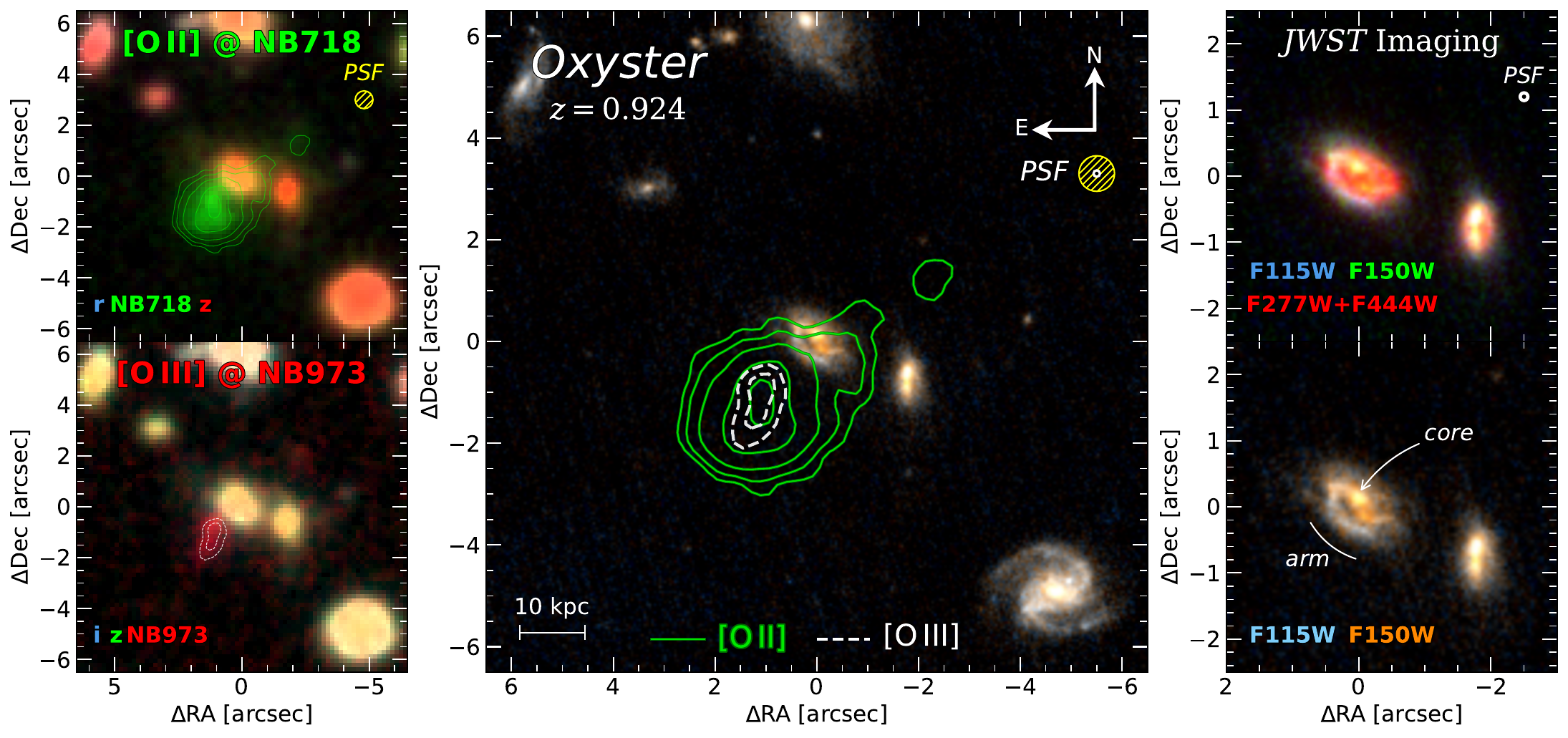}
    \caption{Subaru/HSC and JWST/NIRCam imaging of Oxyster. Left panel: Fake RGB images from Subaru HSC r, i, z, NB718 and NB973 bands (noted in lower left corner of each subpanel) of the emission line nebula with [\ion{O}{2}] marked in green (upper) and [\ion{O}{3}] marked in red (lower) aligned with nebula contours. Middle panel: JWST NIRCam imaging overlaid with [\ion{O}{2}] in green solid line and [\ion{O}{3}] in white dashed line. Right panel: COSMOS-Web JWST/NIRCam fake RGB images of the galaxy. The substructures of the galaxy (core and arm) are annotated. Bands in different colors are demonstrated in the lower left corners. The PSF sizes of HSC and JWST images are shown in the upper right of each panel.
        }
\label{RGBfigs}
\end{figure*}

    Figure~\ref{RGBfigs} summarizes the multiband imaging data for Oxyster and its potential host galaxy. On the HST and JWST images with high spatial resolution, the galaxy displays a red core and a single blue ``arm'', which is the most prominent feature in the JWST/F115W and F150W filters. This peculiar structure and the color contrast between the core and the ``arm'' might contain valuable information related to the galaxy's recent evolution and the potential origin of Oxyster. Therefore, we aim to perform spatially resolved SED fitting to uncover more details, and we will conduct PSF-homogenized aperture photometry to achieve this. 
    
     First, to establish the ``global'' SED of the whole galaxy as a reference, we convolved all images listed in Table~\ref{photodata} to a common Moffat PSF with $\rm{FWHM}=0.8~\rm{arcsec}$ and $\rm{\beta}=2.5$ following the method in \citet{Weaver-2022}. For the HSC/CHORUS images, we run \texttt{SExtractor} \citep{Bertin-1996} and \texttt{PSFEx} \citep{Bertin-2011} to extract the mean PSF model from a $11.8\times11.8~\rm arcmin$ coadd image. For the HST images, we select a bright, unsaturated {\it Gaia} star 18 arcmin from the host galaxy as an empirical PSF model. As for the JWST ones, we rely on the PSF models generated by \texttt{WebbPSF} \citep{Perrin-2012}. A summary of our photometric data to construct the SED is listed in Table \ref{photodata}. After we confirm that all PSF models are well centered through visual inspection, we apply \texttt{pypher} \citep{Boucaud-2016} to generate the PSF-matching kernels and perform the convolution kernel to match the PSFs. \texttt{pypher} accounts for anisotropic features in the PSFs based on Wien filtering, particularly for dealing with JWST images. After the convolution, we select unsaturated stars on the image and measure their growth curves to verify that the PSFs have been homogenized to the design model. Then, we aggressively mask out all objects detected within a 3-$\sigma$ threshold on the image smoothed by a $\sigma=3$ pixel Gaussian kernel using \texttt{photutils} \citep{photoutils} and evaluate the local background. After fitting a Gaussian function to the distribution of flux densities, we confirm the backgrounds are all consistent with zero within 1-$\sigma$ level. Finally, we put down an elliptical aperture with 1.26 arcsec along its major axis, axis ratio 0.75, and position angle $232^{\circ}$. We illustrate this aperture and the host galaxy's SED on the left side of Figure~\ref{SED_all}. We estimate the uncertainties by randomly placing 200 same apertures within 30 arcsec$^2$ of the host galaxy.

    Next, we perform small aperture photometry centered on the core and the ``arm'' of the galaxy. To ensure the spatial resolution is enough, we only rely on the HST F606W \& F814W images and the JWST ones in four bands (marked with an asterisk in Table \ref{photodata}). We apply the same PSF homogenization procedure but use the JWST/F444W image's PSF, which has the lowest resolution, as the reference model. We choose 0.3 arcsec circular apertures for both the core and the ``arm'' by visual inspection. We confirm that a small variation of the central location will not impact any key conclusions from the SED fitting. We show the SEDs of the core \& ``arm'' and highlight these two small apertures on the right panel of Figure~\ref{SED_all}.

    After the aperture photometry, we correct the Galactic dust extinction based on the $A_{\rm{V}}=0.044~\rm{mag}$ value from the \emph{combined15} \citep{Marshall-2006, Green-2015, Drimmel-2003} dust map and assuming the $R_{\rm{V}}=3.1$ Milky Way extinction law.
    
\subsection{Spatially resolved SED fitting} 
    \label{sec:SED}

    Using the galaxy-wide and spatially resolved photometric measurements from the multiband imaging shown in the last section, we perform SED fitting using \texttt{Prospector} \citep{Leja-2019, Johnson-2021} algorithm and the Flexible Stellar Population Synthesis \citep[FSPS,][]{Conroy-2009, Conroy-2010} models. We chose the Kroupa Initial Mass Functions (IMF) \citep{Kroupa-2001}, Padova isochrones \citep{Bertelli-1994, Girardi-2000, Marigo-2008}, and the MILES spectral library \citep{Sanchez-Blazquez-2006} to generate the single stellar population model. We also assume a Calzetti dust attenuation law \cite{Calzetti-2000} and add nebula continuum \& emission lines \citep{Byler-2017} in the fitting. 
    
    More importantly, choosing the star formation history (SFH) model is critical for inferring stellar mass, star formation rate, dust attenuation, and other key SED properties relevant to this work \citep[e.g.,][]{Michalowski-2014, Lower-2020, Mosleh-2025}. In particular, \texttt{Prospector} not only supports ``traditional'' parametric SFH models, such as the delayed-$\tau$ model \citep[e.g.,][]{Sandage-1986}, but also enables SED fitting based on non-parametric SFH models. For instance, the non-parametric SFH model with the ``LogM'' prior investigated in \citep{Leja-2019} assigns SFRs in a series of pre-designed age bins to represent complex but essential events, such as episodes of starbursts, quenching, and rejuvenation, which is particularly appropriate for our spatially resolved SEDs in the core and ``arm'' regions. In this work, we also adopt the same model (the ``LogM'' model). We follow the 8-time-bin design endorsed by \citet{Leja-2019} and rescale the time length to the age at $z=0.924$.
    We also add a separate bin for the recent 30 Myrs to account for potential starburst or quenching events. Table~\ref{Prospect_set} summarizes the key model parameters and the choice of priors. We choose uniform priors within broad boundaries (or semi-uninformative) priors for most parameters except for the gas metallicity, which we fix at the solar value. As shown later, the emission line ratio analysis favors high gas metallicity for Oxyster. And, given that we are only fitting photometric data, the choice of gas metallicity has minimal impact on the results. 

    For the fitting process, we use the dynamic nested sampler \texttt{dynesty} \citep{Speagle-2020} supported by \texttt{Prospector} to explore the high-dimensional parameter space and infer the posterior distributions. We initialize the dynamic nested sampler with 1000 live points (\texttt{nlive\_init}$=1000$) and choose \texttt{dlogz}$=0.01$ to be the stopping criteria, where \texttt{dlogz} represents the remaining prior volume's contribution to the Bayesian evidence. 
    
\begin{figure*}[htbp]
    \centering 
    \includegraphics[width=1\textwidth]{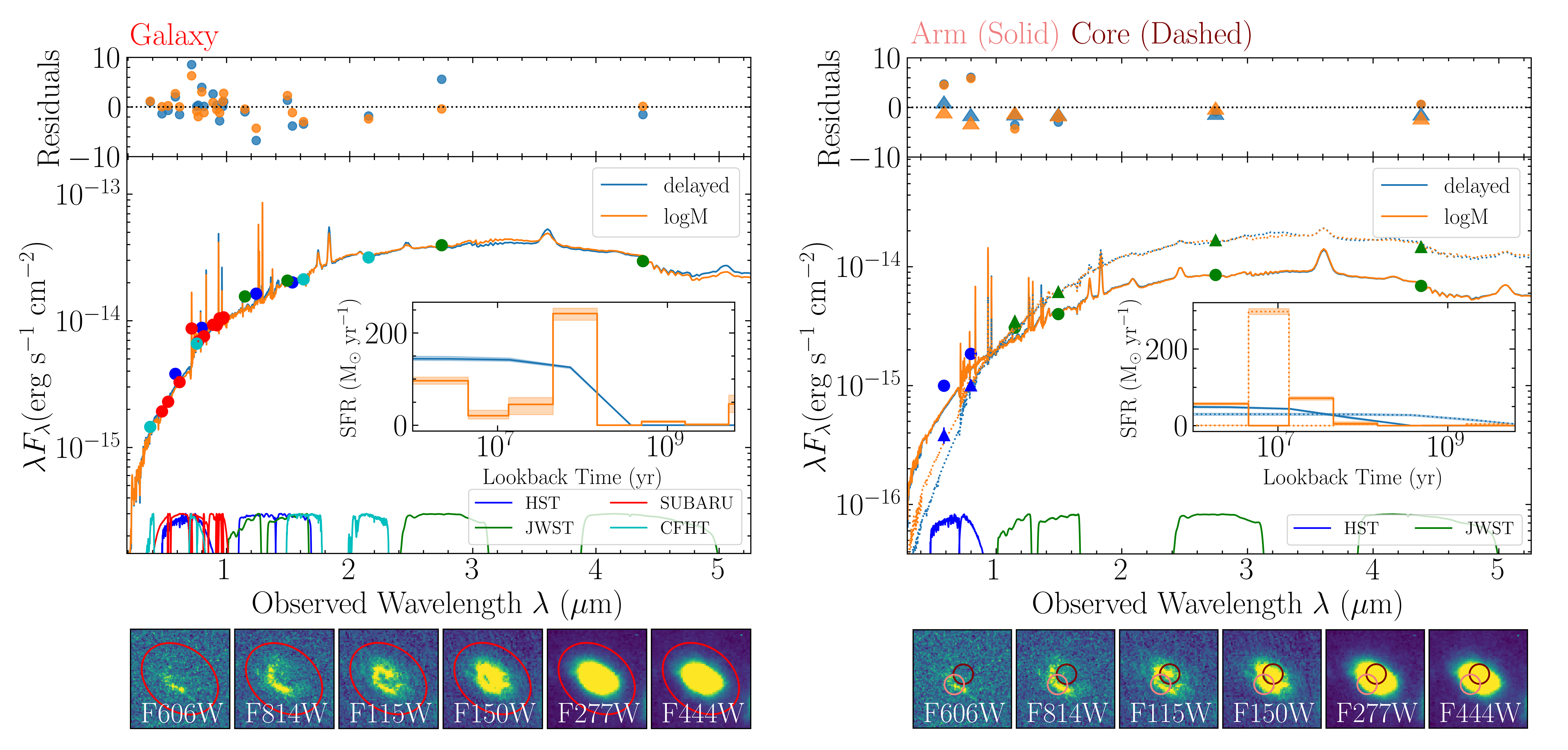} 
    \caption{ The SED of the entire galaxy and substructures. Left Panel: The SED of the entire galaxy. Colored dots represent photometric data from various telescopes, with corresponding filter transmission curves shown below (details in Section \ref{subsec:photometry}). The orange and blue curves indicate the best fits obtained with \texttt{Prospector} for delayed-$\tau$ and logM SFH priors, respectively, with their star formation histories (SFHs) shown in the inset. The top subpanel displays the $\chi^2$ values for the fitting between the observed photometry and the model. HST and JWST cutouts are shown in the lower panel, with the aperture used for the SED analysis marked in red.
    Right Panel: Similar to the left panel but for specific substructures of the galaxy, namely the ``arm" and ``core." Their best-fit spectra are represented by solid and dashed lines, and photometric data are denoted by dots and triangles, corresponding to the pink and dark red apertures shown in the cutouts below.} 
    \label{SED_all} 
\end{figure*}

    \textbf{Given the limited number of photometric bands for the separate structures, we acknowledge that the SED alone provides only modest constraints on complex SFHs. To mitigate the risk of overfitting, we adopted physically motivated priors and a limited number of time bins(8+1 recent bin), following the setup in \cite{Leja-2019}. We further ensured that time bin edges were consistent with expectations for galaxies at $z=0.924$, and that the parameter ranges were broad but not unconstrained.} Table \ref{SED_results} presents the main results of the SED fitting using the stellar mass, SFR, and dust attenuation ($A_V$) values from both the ``$\log$M'' and the delayed-$\tau$ models for the whole galaxy, the core, and the ``arm'' regions. We also show the best-fit SED models and their residuals in Figure \ref{SED_all}. 

\begin{table*}[]
\begin{tabular}{crll}
\hline \hline
\multicolumn{2}{c}{Parameters}                                            & Description                    & Prior/Value                                                                                                                                                                                       \\ \hline \hline
\multicolumn{2}{r}{$z$}                                                   & Spectroscopic redshift          & 0.924                                                                                                                                                                                             \\
\multicolumn{2}{r}{IMF}                                                   & Initial Mass Function          & \cite{Kroupa-2001}                                                                                                                                                                   \\
\multicolumn{2}{r}{$\rm{log}(Z_*/Z_\odot)$}                               & Stellar metallicity            & Uniform: (-2, 0.6)                                                                                                                                                                                \\ \hline
\multirow{2}{*}{Dust Attenuation} & $k'(\lambda)$   & Dust attenuation curve         & \cite{Calzetti-2000}                                                                                                                                                                 \\
                                  & $A_{\rm{V}}$                          & V band attenuation             & Uniform: (0, 6)                                                                                                                                                                                   \\ \hline
\multirow{2}{*}{Nebular Emission} & logU            & Gas-phase ionization parameter & Uniform: (-5, -1)                                                                                                                                                                                 \\
                                  & $Z_{\rm{gas}}$                & Gas-phase metallicity           & $Z_\odot$                                                                                                                                                                                         \\ \hline \hline
\multirow{3}{*}{delayed-tau SFH}  & $t_{\rm{age}}$  & Time since SF began            & Uniform: (0.1, 13.8)                                                                                                                                                                              \\
                                  & $\tau$                                & e-folding timescale            & Uniform: (0.3, 10)                                                                                                                                                                                \\
                                  & $\rm{log}(M_*/M_\odot)$               & Formed stellar mass            & Uniform: (1, 15)                                                                                                                                                                                  \\ \hline
\multirow{3}{*}{``logM" SFH}      & $N_{\rm{bins}}$ & Number of time bins            & 8                                                                                                                                                                                                 \\
                                  & $t_{\rm{bin}}$                        & Lookback time bin edges        & \shortstack[l]{(0, 30 Myr, 100 Myr, 330 Myr, \\ 1.1 Gyr, 3.6 Gyr, 11.7 Gyr, 13.7 Gyr),\\ scaled to cosmic age at z.} \\
                                  & $\rm{log}(M_*/M_\odot)_{\rm{bin}}$    & Formed stellar mass per bin    & Uniform: (5, 11)                                                                                                                                                                                  \\ \hline \hline
\end{tabular}
\caption{Description of the parameter settings of Prospector.}
\label{Prospect_set}
\end{table*}


\subsection{Spectral properties of \emph{Oxyster}}
    \label{subsec:spectra}
\begin{figure*}[htbp]
    \centering
    \includegraphics[width=1\textwidth]{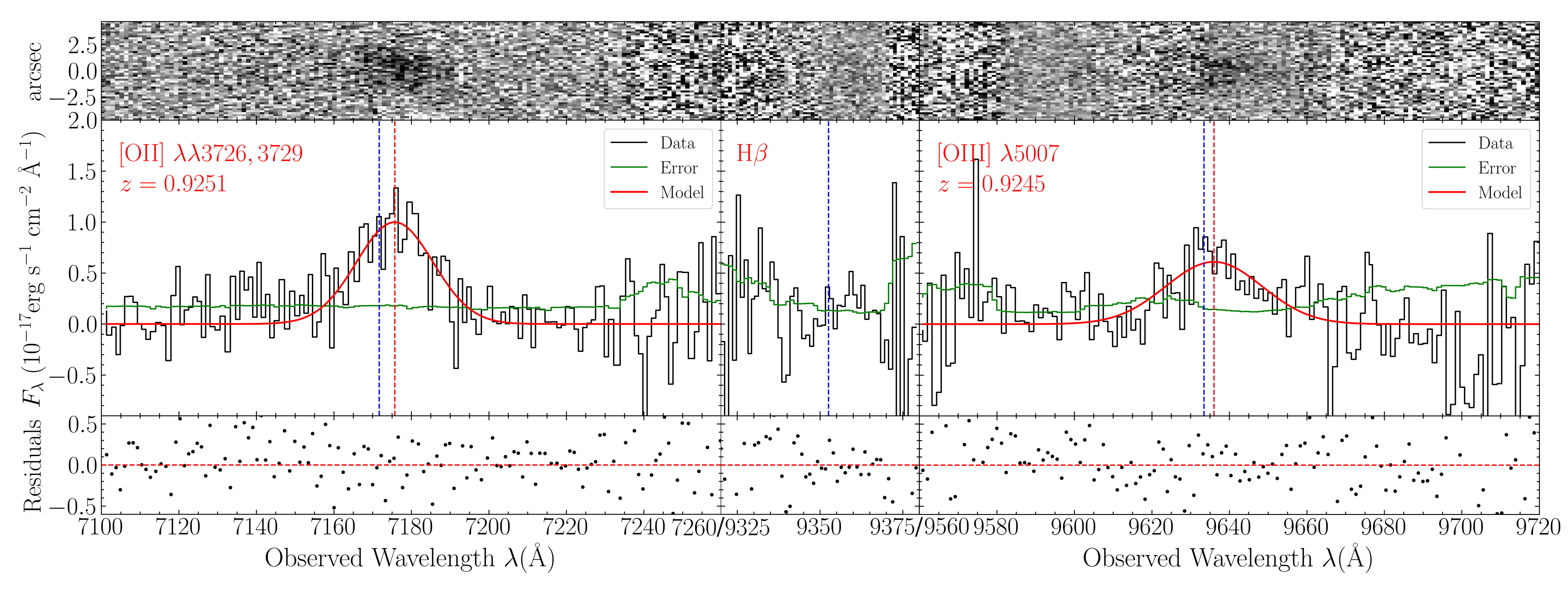}
    \caption{
        Long-slit spectra from Magellan LDSS3, reduced by \texttt{Pypeit}, as described in Section \ref{subsec:spectra}. The upper panel shows the 2D spectra of [\ion{O}{2}], non-detected H$\beta$ and [\ion{O}{3}] separately. The main figures in the middle row represent the 1D spectra (black solid line) and 1-sigma error (green solid line) along with the fitted 1D Gaussian curve (red solid line). Line centers at the host galaxy's redshift($z=0.924$) are demonstrated in vertical blue dashed lines, and the Gaussian centers are shown in red dashed lines. Residuals of the Gaussian fitting are shown in the lower panel with black dots. The horizontal red dashed line corresponds to the zero level. }
\label{O2_spec}
\end{figure*}

    With the reduced spectra (see Section \ref{sec:spectra}), we perform emission line analysis to estimate critical emission line ratios that could help us determine the physical nature of Oxyster. As explained earlier, given that Oxyster was observed as a filler target under a poor seeing condition with a fixed instrument setup that is suboptimal for our purpose, we only achieve significant detections of the {\emph{unresolved}} [\ion{O}{2}] doublet and the [\ion{O}{3}]$\lambda 5007$ line. We measure their integrated fluxes through Gaussian fitting (see Figure \ref{O2_spec}). As for [\ion{O}{3}]$\lambda 4959$, based on the [\ion{O}{3}]$\lambda 5007$ flux, the assumption of a 2.98 intrinsic line ratio, and the local noise level in the spectrum, we expect the non-detection of this weaker [\ion{O}{3}] line. Interestingly, we also do not detect the H$\beta$ Balmer emission line, which is often prominent in emission line nebulae, in the spectrum and can only provide a 3-$\sigma$ lower flux limit of $4.5\times 10^{-18} \rm erg~s^{-1}~cm^{-2}~\r{A}^{-1}$.

    Table \ref{nebula_properties} summarizes the flux, line-of-sight velocity, and the velocity dispersion from the Gaussian fit for the two detected emission line systems. As we do not resolve the [\ion{O}{2}] doublet, we fit them with a single Gaussian function, since we only use the line flux for further analysis. We recommend the readers take the [\ion{O}{2}] velocity and velocity dispersion values with a grain of salt. 

    In addition to the emission line ratio estimation, we will use these properties to calibrate the continuum subtraction from the narrow-band images, which we will describe in the next section.

\begin{table*}[!htbp]
    \centering
    \begin{tabular}{ c c c c c }
    \hline
    \hline
        Structure & SFH model & Mass [$M_{\rm{\odot}}$] & SFR [$M_{\rm{\odot}}$/yr] & $A_{\rm{V}}$ \\ \hline \hline
        Galaxy & $\begin{array}{c} \rm logM \\ \rm delayed~tau \end{array}$&
        $\begin{array}{c} (5.7\pm0.8) \times 10^{10} \\ (1.8\pm0.1) \times 10^{10} \end{array}$ & 
        $\begin{array}{c} 96\pm8 \\ 144\pm5 \end{array}$ & 
        $\begin{array}{c} 2.4 \\ 2.7 \end{array}$ \\ \hline
        ``Arm" & $\begin{array}{c} \rm logM \\ \rm delayed~tau \end{array}$ &
        $\begin{array}{c} (3.5\pm0.5) \times 10^{9} \\ (2.6\pm0.2) \times 10^{9} \end{array}$ & 
        $\begin{array}{c} 57\pm3 \\ 49\pm3 \end{array}$ & 
        $\begin{array}{c} 2.8 \\ 2.9 \end{array}$ \\ \hline
        ``Core" & $\begin{array}{c} \rm logM \\ \rm delayed~tau \end{array}$ &
        $\begin{array}{c} (1.3\pm0.2) \times 10^{10} \\ (3.8\pm0.3) \times 10^{10} \end{array}$ & 
        $\begin{array}{c} 0.6\pm0.5 \\ 30\pm3 \end{array}$ & 
        $\begin{array}{c} 3.7 \\ 3.5 \end{array}$ \\ \hline
    \end{tabular}
\caption{Basic SED results of the galaxy in terms of ``logM'' and ``delayed-tau'' SFH priors. For Mass, SFR, and $A_{\rm{V}}$, the upper values correspond to ``logM'' priors and the lower values to ``delayed-tau'' priors.}
\label{SED_results}
\end{table*}

\begin{table*}[!ht]
    \centering
    \begin{tabular}{c c c c c c c}
    \hline
    \hline
        Emission Line & $\begin{array}{c} \rm Flux \\ \rm [erg/s/\rm cm^2] \end{array}$ &$\begin{array}{c} \rm Luminosity \\ \rm [erg/s] \end{array}$ &  $\begin{array}{c} \mu \\ \rm [erg/s/cm^2/arcsec^2] \end{array}$ & $\begin{array}{c} R \\ \rm [kpc] \end{array}$   & $\begin{array}{c} v \\ \rm [km/s] \end{array}$ & $\begin{array}{c} \sigma \\ \rm [km/s] \end{array}$ \\ \hline \hline
        [O II] $\lambda\lambda$3726,3729 &
        $(2.15 \pm 0.02) \times 10^{-16}$ & $(9.24 \pm 0.09) \times 10^{41}$
        & $(1.76 \pm 0.02) \times 10^{-17}$ & $27 \pm 3$ & -170 $\pm$ 10 & $<$ 475 \\ \hline
        [O III] $\lambda$5007 & 
        $(3.48 \pm 0.22) \times 10^{-17}$ & $(1.50 \pm 0.09) \times 10^{41}$
        & $(4.26 \pm 0.27) \times 10^{-17}$
         &  $21 \pm 3$ & -30 $\pm$ 86 & $<$ 385 \\ \hline
    \end{tabular}
\caption{Basic properties of the nebula. As described in Sec.~\ref{subsec:surface}, flux, luminosity, $\mu$ (surface brightness), and $R$ (radial extent) are derived from narrow-band images following broad-band subtraction and MW dust correction. $v$ (velocity) and $\sigma$ (velocity dispersion) are measured from LDSS3 spectra using 1D Gaussian fitting. Note that the velocity dispersions represent upper limits, and velocities are provided for reference only, given the limited spectral resolution. We did not consider the projection effect.}
\label{nebula_properties}
\end{table*}


\subsection{Surface Brightness Profile of \emph{Oxyster}} 
    \label{subsec:surface}

    While the narrow-band images provide robust evidence of Oxyster's existence, they still contain possible contributions from the continuum emission. To help infer more accurate emission line fluxes and spatial distributions, we construct the emission line surface brightness map of [\ion{O}{2}] ([\ion{O}{3}]) based on the NB718 (NB973) narrow-band images after estimating \& subtracting potential continuum emissions using the nearby broad-band images. 

    First, we estimate the power-law slope ($\beta$ in the $f_\lambda\propto\lambda^\beta$) of the continuum spectrum in the nearby broadband filters based on the best-fit SED from {\tt Prospector}, which is described in Section \ref{sec:SED}. Following the approach proposed in the Appendix of \citet{Li2024ApJS..275...27L}, we derive an emission line map based on the empirical estimator:

    \begin{equation}
        F_{\rm line} = \left(\epsilon_{\rm line, NB} \langle f_{\nu} \rangle_\mathrm{NB} - \sum_i\alpha_{BBi}\epsilon_{\rm line, BBi} \langle f_{\nu} \rangle_\mathrm{BBi} \right) \frac{c}{\lambda_{\rm line}^2}.
    \end{equation}

    In this formula, $F_{\rm line}$ is the estimated line flux; $\langle f_{\nu} \rangle_\mathrm{NB}$ is the narrow-band flux density covering the emission line; $\langle f_{\nu} \rangle_\mathrm{BBi}$ is the multiple broad-band flux density used to estimate the continuum; $\epsilon_{\rm line, NB}$ and $\epsilon_{\rm line, BB}$ are calibrated effective bandwidths, which are effective bandwidths considering the continuum slope $\beta$ and the off-pivot wavelength of the emission line.
    And, $\alpha_i$ is the weight with $\sum_i\alpha_{BBi} = 1$; $c$ is the light speed, and $\lambda_{\rm line}$ is the observed wavelength of the emission line.
    We refer to the Appendix in \citet{Li2024ApJS..275...27L} for details of this method and the definition of $\epsilon_{\rm line, NB}$ and $\epsilon_{\rm line, BB}$.

    For the [\ion{O}{3}] line in the NB973 band, we took $z$-band as the reference broad-band; hence the formula becomes:

    \begin{equation}
        F_{\rm [OIII]} = \left(\epsilon_{\rm [OIII],NB973} \langle f_{\nu} \rangle_\mathrm{NB973} - \epsilon_{\rm [OIII], z} \langle f_{\nu} \rangle_\mathrm{z} \right) \frac{c}{\lambda_{\rm [OIII]}^2},
    \end{equation}
    where $\epsilon_{\rm [OIII],NB973} = 264.31\AA$ and $\epsilon_{\rm [OIII],z} = 319.64\AA$.
    
    For the [\ion{O}{2}] line in the NB718 band, we took $r$-band and $z$-band to estimate the continuum with an equal weight of $\alpha_r=\alpha_z=\frac{1}{2}$, hence the formula can be written as:

    \begin{equation}
    \begin{aligned}
        F_{\rm [OII]} &= \left(\epsilon_{\rm [OII],NB718} \langle f_{\nu} \rangle_\mathrm{NB718}\right.\\
        &\left.- \frac{1}{2}\epsilon_{\rm [OII], r} \langle f_{\nu} \rangle_\mathrm{r} - \frac{1}{2}\epsilon_{\rm [OII], z} \langle f_{\nu} \rangle_\mathrm{z} \right) \frac{c}{\lambda_{\rm [OIII]}^2},
    \end{aligned}
    \end{equation}
    where $\epsilon_{\rm [OII],NB718} = 109.67\AA$, $\epsilon_{\rm [OII],r} = 161.24\AA$, and $\epsilon_{\rm [OII],z} = 67.17\AA$.

\begin{figure*}[htbp]
    \centering
    \includegraphics[width=1\textwidth]{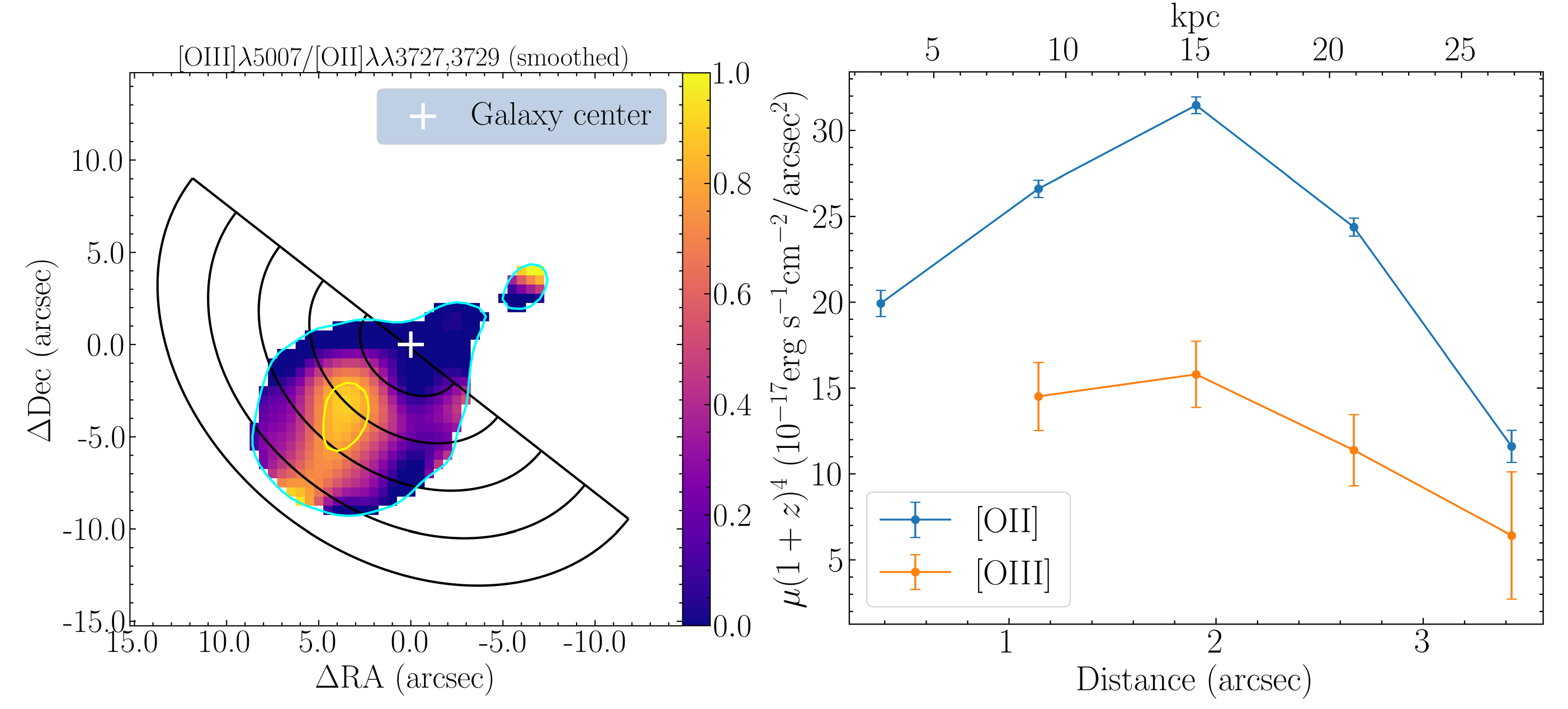}
    \caption{2D line ratio map and 1D surface brightness radial profile of Oxyster.
        Left panel: Smoothed [\ion{O}{3}]$\lambda 5007$/[\ion{O}{2}]$\lambda\lambda 3726,3729$ map from the narrow bands subtraction described in Section \ref{subsec:surface}. Cyan and yellow solid lines indicate [\ion{O}{2}] and [\ion{O}{3}] contours that are 2-sigma above the background after Gaussian smoothing. Elliptical annuluses are shown in black solid lines. Right panel: Radial profiles of two emission lines. The surface brightness $\mu$ is measured inside the intersection between the [\ion{O}{2}] nebula contours and the annulus regions. The ``distance'' refers to each bin's middle point along the elliptical annulus's minor axis.}
    \label{radialprofile} 
\end{figure*}

    Because the PSFs among these bands are very similar, we retain the original images without matching PSFs and follow these formulae to obtain the emission maps. We present emission line maps in Figure \ref{radialprofile}(a), showing that the [\ion{O}{2}] emission extends up to $\sim30$ kpc from the host galaxy. The cyan (yellow) contour outlines the 2$\sigma$ threshold of the [\ion{O}{2}] ([\ion{O}{3}]) emissions after a $\sigma=2$ pixel Gaussian kernel smoothing. We then calculate each emission line's ``total'' luminosities by integrating within each contour after correcting the cosmological dimming effect at $z=0.924$. 

    Next, we estimate the radial surface brightness profiles of [\ion{O}{2}] and [\ion{O}{3}] emission lines using five elliptical radial bins defined based on the aperture of Oxyster's host galaxy. Although this elliptical annulus does not follow the contour of Oxyster, we choose this option to depict the emission line intensity variation relative to the host galaxy, which could drive the creation of Oxyster. Each annulus has a 4-pixel bin width, matching the FWHM of the HSC PSF to avoid oversampling the radial variation. Within each annulus, we use the region above the 2$\sigma$ threshold of [\ion{O}{2}] map to estimate the mean emission line surface brightness to keep the integral area the same for both emission lines. We show the results on the right panel of Figure \ref{radialprofile}.

    As we can see, Oxyster's [\ion{O}{2}] emission is more extended than the [\ion{O}{3}] and can be seen as far as $\sim 25$ kpc away from the host galaxy when the [\ion{O}{3}] emission only reaches $\sim 15$ kpc. In the radial direction, both the [\ion{O}{2}] and [\ion{O}{3}] emissions peak between 10-15 kpc. Meanwhile, along the azimuthal direction, both emission lines are the strongest along the direction that is always perpendicular to the major axis of the host galaxy. Oxyster appears to be a single ``blob'' in these emission lines. However, it is difficult to comment further on the nebula's morphology or structure, given the narrow-band image's spatial resolution. On the other side of the host galaxy, the NB718 image also reveals a tiny cloud of [\ion{O}{2}] emission at a $\sim 20$ kpc distance. Unfortunately, the current data quality cannot verify its physical connection to the primary emission line region. 

    More importantly, the most notable character of Oxyster from the emission line maps is not the spatial distribution but the low [\ion{O}{3}] to [\ion{O}{2}] emission line flux ratio (or O32=$f_{\text{[OIII]}\lambda5007}/f_{\text{[OII]}\lambda\lambda3276,3279}$), which is often used to indicate the hardness of the ionizing radiation field. The color map on the left side of Figure \ref{radialprofile} indicates the spatially resolved O32 ratio, which is $<1$ over the entire nebula region. While the [\ion{O}{3}] surface brightness is below the 2$\sigma$ limits for a significant fraction of the [\ion{O}{2}] nebula, making the per-pixel O32 ratio more uncertain, when averaged within the elliptical annulus using the same footprint, the radial profiles of the two emission lines also reflect the low O32 ratio, providing an essential constraint about Oxyster's origin. We summarize the properties of the emission lines in Table \ref{nebula_properties}.

\section{Results}
    \label{sec:results}

    In this work, we report the serendipitous discovery of an extended emission line nebula associated with a star-forming galaxy at redshift $z=0.924$, identified through the [\ion{O}{2}] and [\ion{O}{3}] emission lines using narrow-band filters NB718 and NB973 from the Subaru/HSC imaging in the COSMOS field. We perform multi-band photometric measurements of the galaxy and the nebula, which help us estimate stellar mass, star formation rate, and dust attenuation of different components of the host galaxy and extract radial profiles of the emission line intensities on the nebula. We also confirm the redshift of the nebula and its physical connection to the host galaxy using spectroscopic observation. Below, we summarize the key findings of the nebula and the host galaxy.

    On the emission nebula itself, we find that Oxyster extends out to $\sim30$ kpc in [\ion{O}{2}] and $\sim20$ kpc in [\ion{O}{3}] emission line, indicating extended circumgalactic gas. The projected angular dimension of the nebula is larger or at least comparable to the host galaxy. Both the [\ion{O}{2}] and [\ion{O}{3}] surface brightness distributions peak at $\sim 15$ kpc from the center of the host. The [\ion{O}{2}] and [\ion{O}{3}] luminosities are about $1.2 \times 10^{42}~\rm{erg/s}$ and $3.1 \times~10^{41}~\rm{erg/s}$. The [\ion{O}{3}] luminosity of Oxyster is comparable to extended narrow-line regions around low-redshift AGNs, including the famous Hanny's Voorwerp, which is believed to be ionized by a fading AGN \citep{Schawinski-2010}. The spatially resolved O32 ratio shows the distribution peaking at the center of Oxyster and decreasing toward the outside. Interestingly, the O32 ratio is below 1.0 over the entire nebula, suggesting a relatively low ionization state.

    Using a quick low-resolution Magellan/LDSS3 spectrum through the nebula, we confirm the redshift of the nebula through the detection of the (unresolved) [\ion{O}{2}] doublet and the [\ion{O}{3}]$\lambda 5007$ emission line. Interestingly, we did \emph{not} detect the $\rm{H\beta}$ emission line. While the spectrum confirms the $<1.0$ O32 ratio and its low-ionization implication, the lower limit of the [\ion{O}{3}] to $\rm{H\beta}$ ratio seems to point in the other direction. We will discuss the possible origin of this unique feature of ``Oxyster'' later. The relative line-of-sight velocities of the [\ion{O}{2}] and [\ion{O}{3}] emission lines are $170 ~\rm{km/s}$ and $30 ~\rm{km/s}$, slightly redshifted from the host galaxy. The upper limits of velocity dispersion are $475 ~\rm{km/s}$ and $385 ~\rm{km/s}$ for [\ion{O}{2}] and [\ion{O}{3}], respectively.

    As for the host galaxy of Oxyster, the most prominent feature is a distinctive one-sided ``spiral arm''-like structure in the {\it HST}/ACS F606W and F814W images (rest-frame $\sim315$ and $\sim 423$ nm) that outshines the entire galaxy. Interestingly, the arm is located on the same side of the host galaxy as the ``Oxyster'' nebula. Through global and spatially resolved SED fitting with different star formation history priors, we find that the host galaxy is a moderate starburst galaxy with a total stellar mass of $\sim 2$-$6\times 10^{10}~\rm{M_\odot}$ and a star formation rate of $\sim 100$-$150\ M_{\odot}/{\rm yr}$. Based on the best-fit SFH, both the delayed-$\tau$ and non-parametric ($\log M$) models suggest the host galaxy experienced the onset of a starburst starting from $\sim 10^8$ years ago. The core of the galaxy is more massive and mature than the arm, with a stellar mass of $\sim 1.5$-$4\times 10^{10}~\rm{M_\odot}$ and a lower star formation rate of $\sim 1$-$30\ M_{\odot}/{\rm yr}$. On the other hand, while the arm region only contributes to about $1/10$ of the stellar mass of the galaxy ($\sim 3$-$4\times 10^{9}~\rm{M_\odot}$), its $\sim 50\ M_{\odot}/{\rm yr}$ SFR suggests a much higher specific star formation rate and a younger stellar population, which is supported by its brightness rest-frame UV to optical-blue bands (e.g., HST/F606W, F814W) that are dominated by young stars. The best-fit SFH model of the arm suggests a starburst intensified in the last $10^7$ years. The arm is also less dust-obscured than the core, with $A_V \sim 2.8$ mag compared to $A_V \sim 3.7$ mag. The best-fit SED model sets a 5\% upper limit for AGN contribution. The VLA 3GHz catalog classified the host galaxy as a mid-to-low luminosity AGN based on the radio excess relative to the expected value supported by an IR-based SFR value. 
    \textbf{However, our SED fitting yields an SFR about twice as high, which reduces the implied radio excess.} In either case, the AGN activity of the host galaxy is not prominent. It is worth noting that the host galaxy also has a physical companion at redshift $z=0.922$ with a projected separation of 80 kpc, which may have interacted with the host galaxy.

    In summary, the ``normal'' nature of the host galaxy makes ``Oxyster'' an intriguing example of a potentially large population of extended emission line nebulae around the star-forming system at intermediate redshift, where the cosmic star formation rate density is higher than today. The fact that a nebula like ``Oxyster'' could be identified in narrow-band imaging surveys might show us a possible approach to searching for them systematically soon, providing us with a valuable dataset to explore the CGM. Meanwhile, ``Oxyster'' does possess some unique features, such as the low O32 ratio, the ``missing'' Balmer emission lines, and the asymmetric morphology of the host galaxy, all of which beg for further discussion. 

\section{Discussion}
    \label{sec:discuss}

    So far, we have established that 1. Oxyster is a giant emission line nebula around a star-forming galaxy, much more extended than typical extended narrow-line regions around low-redshift AGNs; 2. While being above the $z\sim 1$ star-forming main sequence and showing evidence of a possible recent merger event, the host galaxy does not host a strong AGN; 3. While the lower limit of observed [\ion{O}{3}]$\lambda 5007$/H$\beta$ suggests high ionization that typically originates from AGN, [\ion{O}{3}]$\lambda 5007$/[\ion{O}{2}]$\lambda\lambda 3726,3729$ ratio (short for O32 ratio below) is lower than 1.0 for the whole nebula, which is unusual for AGN-ionized gas with high excitation levels. In this section, we discuss the possible origins of Oxyster, the ionization mechanism of the nebula, and the implications of the host galaxy's properties.

\subsection{Excitation Mechanism of Oxyster}

\subsubsection{AGN Ionization} 
    \label{sec:AGN-ionize}
    
\begin{figure}[htbp]
    \centering
    \includegraphics[width=0.47\textwidth]{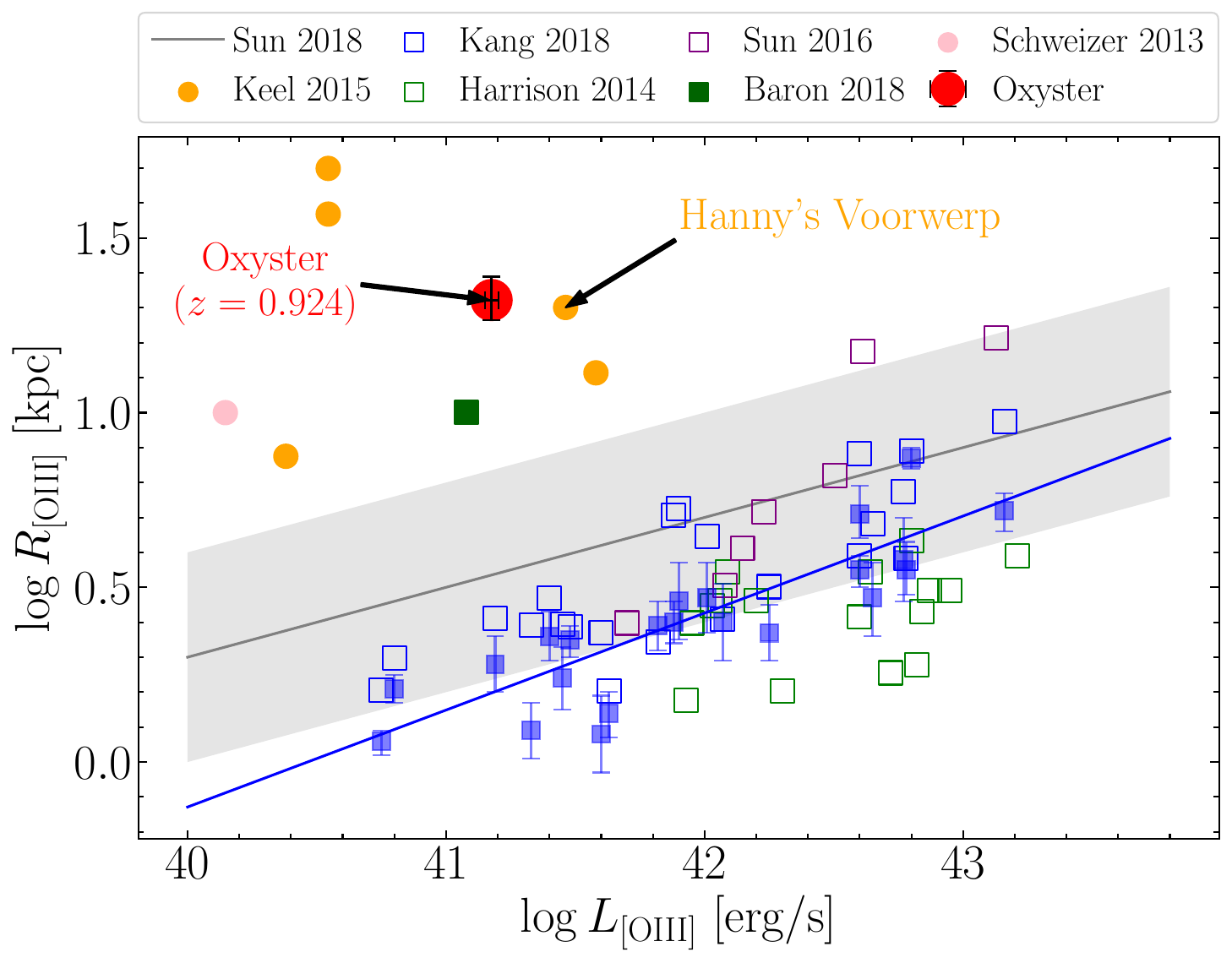}
    \caption{
        Size and luminosity of extended [\ion{O}{3}] emission line regions around (fading) AGNs measured from relative works: \cite{Sun-2018}, \cite{Kang-2018}, \cite{Sun-2017}, \cite{Schweizer-2013}, \cite{Keel-2015}, \cite{Baron-2018}. Nebulae with active AGN hosts are marked in squares, while fading AGN hosts are in circles. The gray solid line and the gray region represent the size-luminosity relation and the scattering for extended emission-line regions of obscured AGN from samples in \cite{Sun-2018}. The blue solid line represents the relation described in \cite{Kang-2018}. The blue solid squares refer to the kinematic sizes and open squares to the photoionized sizes of the former samples. The red dot represents \emph{Oxyster}, which is close to the famous Hanny's Voorwerp.
    } 
    \label{param_space}
    \end{figure}

    At low redshift, extended narrow emission line nebulae traced by [\ion{O}{2}] or [\ion{O}{3}] emissions are often associated with ongoing or past AGN activities. While we find no direct evidence of a luminous AGN in Oxyster, we compare it with other emission line nebulae from the literature on the [\ion{O}{3}] luminosity and size relation (Figure \ref{param_space}), where ``size'' means the maximum extent of the emission, not the diameter of a nebula itself. 
    Firstly, we show the locations of extended narrow-line regions (ENLR; e.g., \citealt{Harrison-2014, Kang-2018, Sun-2017}). These ENLRs represent the ionized outflowing gas from active AGNs and form a clear luminosity-size relation, as a brighter AGN can photoionize a larger region of its CGM. Oxyster's [\ion{O}{3}] luminosity is at the lower end of this sequence. More importantly, Oxyster is much larger at a given [\ion{O}{3}] luminosity, highlighting its difference from this population. As mentioned, our SED fitting of the host galaxy reveals no clear evidence of a luminous AGN, and its radio continuum emission could be accounted for by the starburst. Based on the upper limit of the Chandra data and assuming a $\Gamma = 1.8$ photon index for K-correction, we derive the X-ray luminosity of Oxyster's host is $\log(L_{\rm{X}}/\rm{erg~s^{-1}}) < 41.9$, making it well below the criteria for an ``X-ray AGN'' \citep[e.g.,][]{Fotopoulou-2016}. While we cannot rule out the possibility of a very low luminosity or obscured AGN at the host's center, it is clear that Oxyster does not fit the ENLR description, which is also consistent with the observed low O32 ratio. 
    
    Interestingly, \cite{Baron-2018} reported the discovery of an outflowing ionizing gas nebula out to 17 kpc around an E$+$A post-starburst galaxy at $z=0.12$ that also hosts an AGN with a $\sim 10^{44}$ erg/s bolometric luminosity (see also, e.g., \citealt{Baron-2017})
    This system has a very similar [\ion{O}{3}] luminosity to Oxyster. They inferred that the galaxy has experienced a starburst lasting for $\sim 400$ Myrs, with a peak SFR ($\sim 120\ M_{\odot}/\rm yr$) similar to Oxyster's host, and is quenched by the AGN feedback in $\sim 60$ Myrs after $\sim 10^9 M_{\odot}$ of gas has been expelled. While the nebula's size is above the luminosity-size relation for ENLRs, it is still smaller than Oxyster's. 
    
    Meanwhile, extended emission line nebulae are found around ``dead'' AGNs. As mentioned, the poster child of low-redshift emission line nebula, ``Hanny's Voorwerp'', is believed to be the ``light echoes'' of a QSO in the past and is intriguingly similar to Oxyster on the luminosity-size plane, reaching $>20$ kpc away from the galaxy (Figure \ref{param_space}). \cite{Keel-2015} also discovered several ``Hanny's Voorwerp'' analogs in the local Universe, two of which are much larger than Oxyster with lower [\ion{O}{3}] luminosities. They suggested that these cases represent the rapidly fading AGNs, whose ionizing luminosity dropped $>100$ fold in the timescale of several $\times 10^5$ years. Similarly, \cite{Schweizer-2013} presented an even fainter, 10 kpc [\ion{O}{3}] nebula next to NGC 7252 with no detection of AGN activity in the host galaxy. While these nebulae around fading AGNs occupy roughly the same parameter space on the luminosity-size plane around Oxyster, they are characterized by a higher O32 ratio, consistent with being photoionized by AGN's hard radiation. This crucial difference makes it unclear whether we can use the ``fading AGN'' scenario to explain Oxyster.

    To quantitatively examine the possibility of AGN photoionization, we model Oxyster's emission line ratios (or lower limit) using the \texttt{CLOUDY} \citep{Ferland-1998, Chatzikos-2023} photoionization framework. 
    To estimate the H$\beta$ upper limit from the LDSS-3 spectra, we assumed the same Gaussian profile as [\ion{O}{2}] and scaled the amplitude to three times the RMS noise level at the corresponding wavelength, yielding a lower limit of [\ion{O}{3}]$\lambda5007$/H$\beta$ = 1.7. We place a gas cloud 10 kpc away from the ionizing source with the open geometry configuration in \texttt{CLOUDY}. Following \citet{Baron-2019}, we explored four AGN models as the ionizing sources with a mean photon energy of 2.56 Ryd, 2.65 Ryd, 3.15 Ryd, and 4.17 Ryd. Since [\ion{O}{2}] and [\ion{O}{3}] are collisionally excited lines whose flux depends exponentially on the electron temperature, we expect negligible contribution to the line flux from $T < 4000$ K gas. The radiation propagation stops when the gas temperature is lower than 4000 K, or the radiation travels 10 kpc in the cloud. Given the low O32 ratio, we focused on the ionization parameter range -2 to -4. We explored a wide metallicity range from 0.2 to 2.0 $\times \ Z_\odot$. We also varied the gas density of the cloud to match Oxyster's observed properties. When the hydrogen density is too high (e.g., $\sim 10~\rm{cm}^{-3}$), all ionizing radiation will be absorbed in a thin slab, inconsistent with the 10 kpc size of Oxyster. On the other hand, when the density becomes too low (e.g., $\sim 0.1 ~\rm{cm}^{-3}$), the resulting line emission is too weak as the line emission scales with density$^2$). Therefore, we adopt the gas density of $n \sim 1~ \rm{cm}^{-3}$ that generally matches the observed size and luminosities of the nebula.

\begin{figure*}[htbp]
        \centering 
        \includegraphics[width=0.7\textwidth]{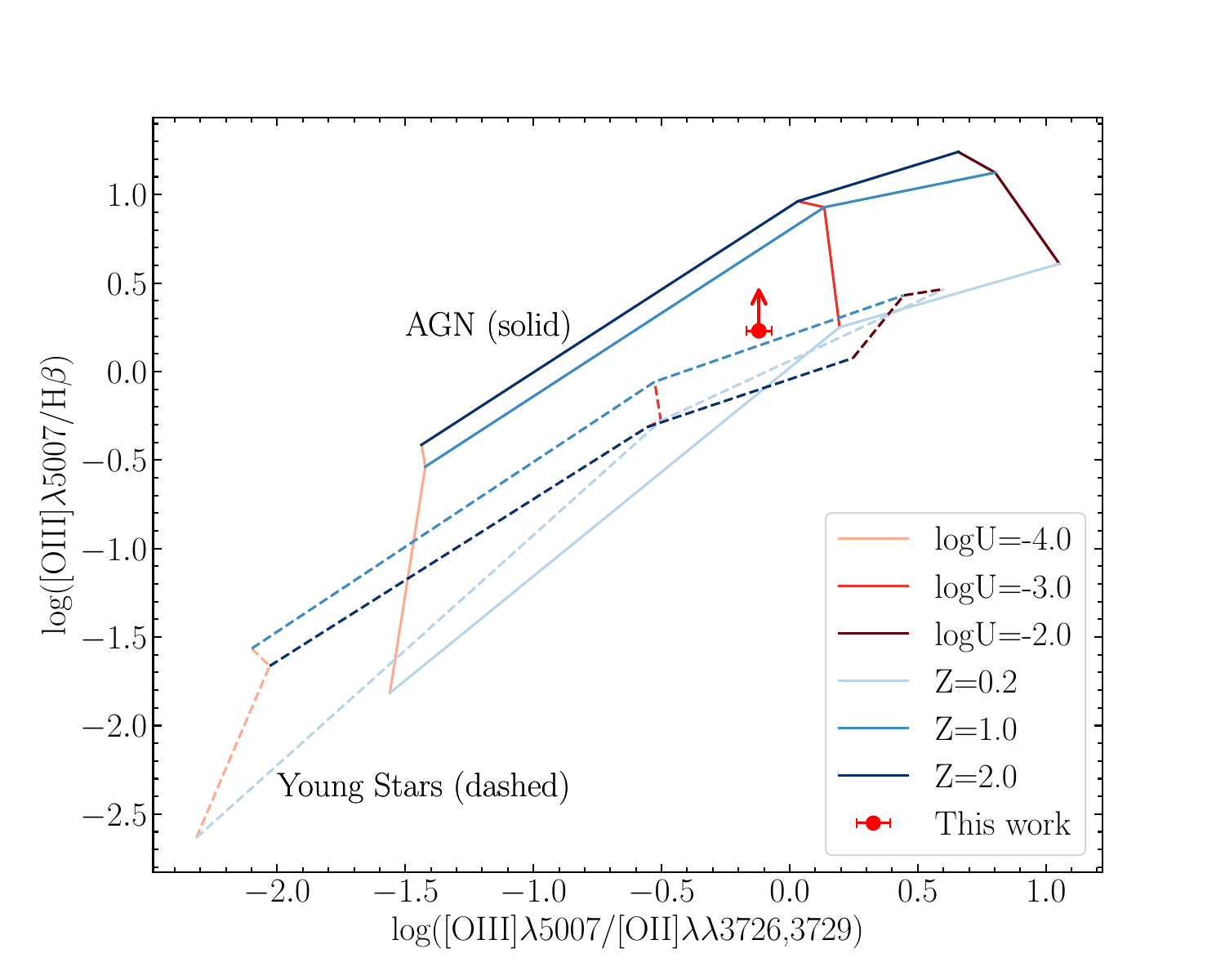}
        \caption{
            Photoionization grids of line ratios from \texttt{CLOUDY} with AGN SED shape (shown as solid lines, from \cite{Baron-2019}) and stellar SED shape (shown as dashed lines, from the stellar spectrum of constant SFH in recent 100 Myr using \texttt{FSPS}) for $n_{\rm{H}} = 1~\rm{cm^{-3}}$. The spectrum O32 ratio (the red dot and arrow) suggests an AGN origin with $\log$U below -3.
            } 
    \label{photogrids}
    \end{figure*}

    Figure \ref{photogrids} compares Oxyster's position on the O32 and [\ion{O}{3}]/H$\beta$ ratio space with the \texttt{CLOUDY} model grid. With only an upper limit of the [\ion{O}{3}]/H$\beta$ ratio, the model can broadly constrain the ionization parameter to $\log U < -3$ and gas metallicity at $>0.2 \times Z_\odot$. 
    
    Following the empirical relation between the ionization parameter and AGN luminosity \citep{Baron-2019}: 

    \begin{equation}
        n_{\rm e} \approx 3.2 \left( \frac{L_{\rm bol}}{10^{45}~\rm erg/s} \right) \left( \frac{r}{1~\rm kpc} \right)^{-2} \left( \frac{1}{U} \right) \rm cm^{-3}
    \end{equation}
    
    \noindent, we estimate the bolometric luminosity of the AGN to be $\sim 1\times10^{42}$ to $\sim 1\times10^{43}$ erg/s, consistent with a very low luminosity AGN. This corresponds to $L_{\rm{X}} = 1\times10^{41} \sim 1\times10^{42}$ erg/s assuming the ratio between bolometric luminosities and hard X-ray luminosities $K_{\rm{X}} = 10$ (e.g., \citealt{Duras-2020}). 
    
    The X-ray luminosity is comparable to the upper limit derived from the Chandra detection limit. Although Oxyster shares a similar [\ion{O}{3}] luminosity and nebula size with Hanny's Voorwerp, we could not borrow the ``fading AGN'' picture where the ionizing flux abruptly drops in the time scale of $10^5$ yr (based on the light speed and the largest radii of the nebula). This suggests that Oxyster is more consistent with a low-luminosity AGN sustaining the ionization of a low-excitation nebula with a low O32 ratio than a rapidly fading AGN. \textbf{Moreover, the spatial variation of the O32 ratio appears inconsistent with expectations from a fading AGN or fossil nebula scenario, where [\ion{O}{3}] should disappear more rapidly as distance increases \citep{Binette-1987}. The persistence of strong [\ion{O}{3}] in central regions of higher inferred density may point instead to many small structures in the nebula center, or directionally-escaped radiation.}

    Although the model grid seems to provide a feasible AGN ionization model for Oxyster, it dramatically under-predicts ($\sim 10 \times$) the [\ion{O}{2}] and [\ion{O}{3}] luminosities. \textbf{While Oxyster does not appear to host a powerful AGN, models involving cosmic-ray (CR) heating have been shown to enhance [\ion{O}{2}] emission in extended nebulae \citep[e.g.,][]{Gagne-2014}. Although the low radio luminosity of the host galaxy makes strong CR influence unlikely, we note that such models could be worth exploring in future work.} Taken all together, the AGN models we explored struggle to explain Oxyster. We will turn our attention to photoionization by starburst in the next section.

\subsubsection{Photoionization by Starburst}

\begin{figure}[htbp]
        \centering 
        \includegraphics[width=0.47\textwidth]{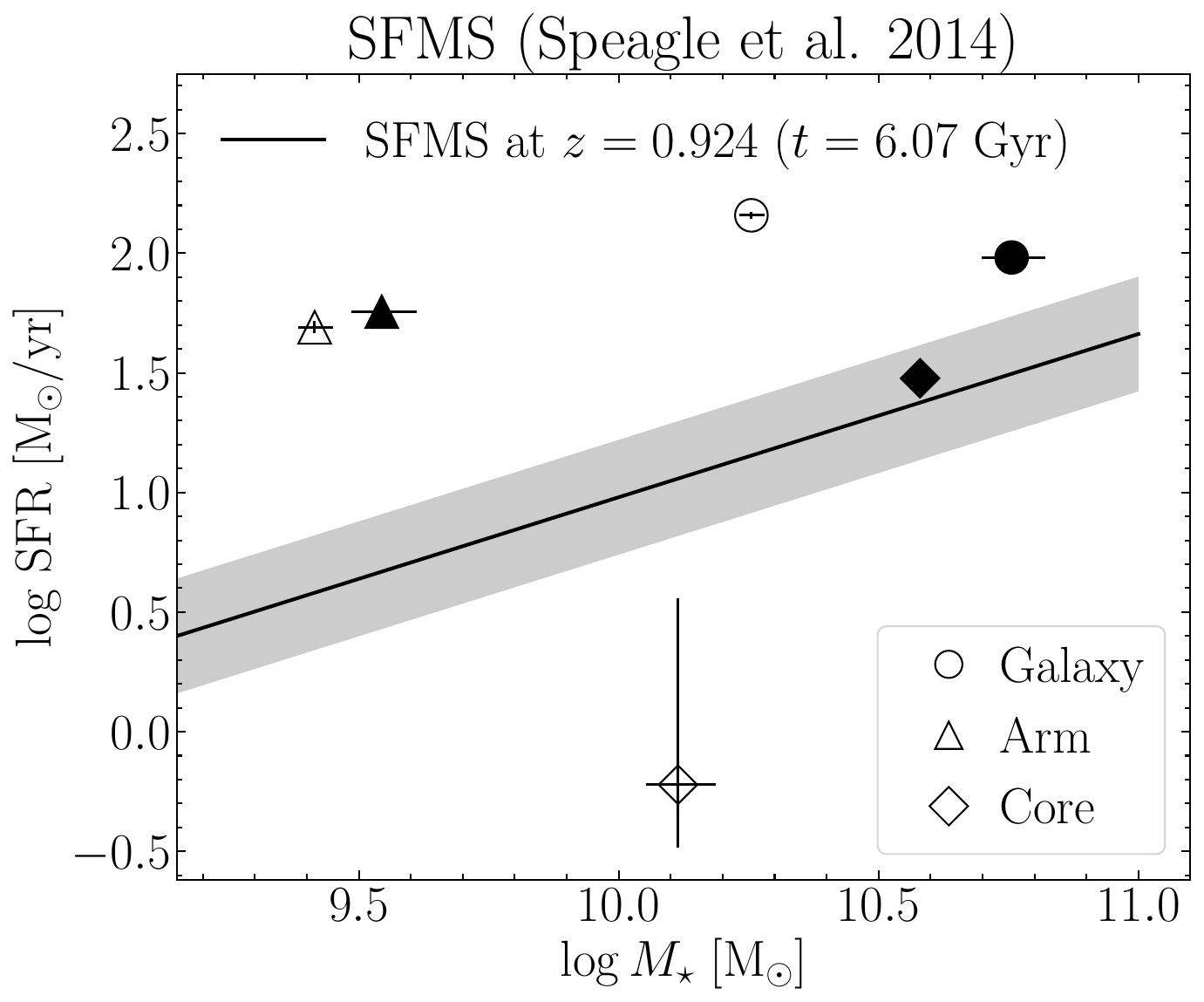}
        \caption{
            The SFR and mass from SED fitting results related to the star-forming main sequence (SFMS) from \cite{Speagle-2014}. The solid black line and gray shaded region represent the SFMS and intrinsic scattering at the cosmic age of $z=0.924$. The circle, triangle, and diamond points stand for different components from the SED fitting: Galaxy, Arm, and Core. The solid symbols represent results from the SFH model with the least total chi-square: logM for Galaxy and Arm, delayed-tau for Core. The open symbols are from the previous SFH. This figure illustrates that the galaxy and the ``arm'' are far above the SFMS and experiencing a starburst.} 
    \label{SFMS}
    \end{figure}

    Given the SED fitting results, Oxyster's host galaxy has been experiencing a starburst for $\sim 10^8$ yrs with its peak SFR $>200 M_{\odot}/{\rm yr}$ (see Figure \ref{SFMS}). In the ``arm'' region, a more recent starburst event might be triggered from $\sim 10^7$ years ago. Both time scales are at least two orders of magnitude longer than the light-traveling time from the galaxy center to the ionization edge of Oxyster. Therefore, photoionization by starburst is a potential scenario worth exploring. Following the same cloud geometric setup for the AGN \texttt{CLOUDY} model, we set the ionizing source as a young starburst with a constant star formation rate for 100 Myrs. We use the {\tt FSPS} model with a Kroupa IMF \citep{Kroupa-2001} and solar metallicity to generate the SED. 

    In Figure \ref{photogrids}, the starburst model grid offsets from the AGN model. While it has no trouble explaining the O32 ratio with an ionizing parameter $-3.0<\log U<-2.0$, the lower limit of the [\ion{O}{3}]/H$\beta$ ratio presents a challenge as the model moderately underpredicts the ratio. However, due to the oversimplified nature of our toy model and the small parameter space explored, we believe the starburst model is still interesting, especially since the starburst model can explain the observed [\ion{O}{2}] and [\ion{O}{3}] luminosities. At the fixed gas density of $n_{\rm{H}}=1~\rm{cm}^{-3}$, a starburst model with an ionization parameter of $\log U=-2.6$, a gas metallicity of $Z=0.2 \times Z_{\odot}$ can reproduce the $> 1\times10^{41}$ erg/s [\ion{O}{3}] luminosity. Assuming that $\sim 10$\% of the ionizing radiation from young stars can escape and leak into the CGM \citep[][]{Inoue-2006, Leitet-2013}, the \cite{Kennicutt-1998} relation predicts a $\sim 97 M_{\odot}/{\rm yr}$ SFR, which is close to the value derived for Oxyster's host galaxy through SED fitting. And, unlike the ``low-luminosity AGN'' hypothesis, we do have clear evidence of an ongoing starburst in the host galaxy (especially in the ``arm'' region) and its inferred SFH to support the starburst scenario where the young stellar population contributed to the photoionization of Oxyster for $>10^5$ years.

    However, instead of concluding that Oxyster is a nebula powered by a recent starburst, we remind the readers that the lack of detections of the Balmer emission lines (especially the H$\beta$ line), the low-velocity resolution of our Magellan spectrum, and the oversimplified modeling approach all leave room for further investigation. 
    
    In summary, while a low-luminosity AGN could explain the high [\ion{O}{3}]$\lambda$5007/H$\beta$ ratio, it struggles to reproduce the observed [\ion{O}{2}] emission line luminosities. On the other hand, a starburst model could provide sufficient ionizing photons to illuminate Oxyster to reach roughly the observed luminosity, but the limited parameter space we explored in \texttt{Cloudy} fails to reach the lower limit of [\ion{O}{3}]$\lambda$5007/H$\beta$. Based on these, a hybrid scenario where both low-luminosity AGN and starburst contribute to the photoionization of Oxyster is a plausible and interesting scenario to explore. Eventually, deeper spatially resolved observations, e.g., with an IFU, are required to resolve additional emission lines that can better constrain the ionizing radiation source.

\subsubsection{Ionization by Shocks}

\begin{figure}[htbp]
        \centering
        \includegraphics[width=0.47\textwidth]{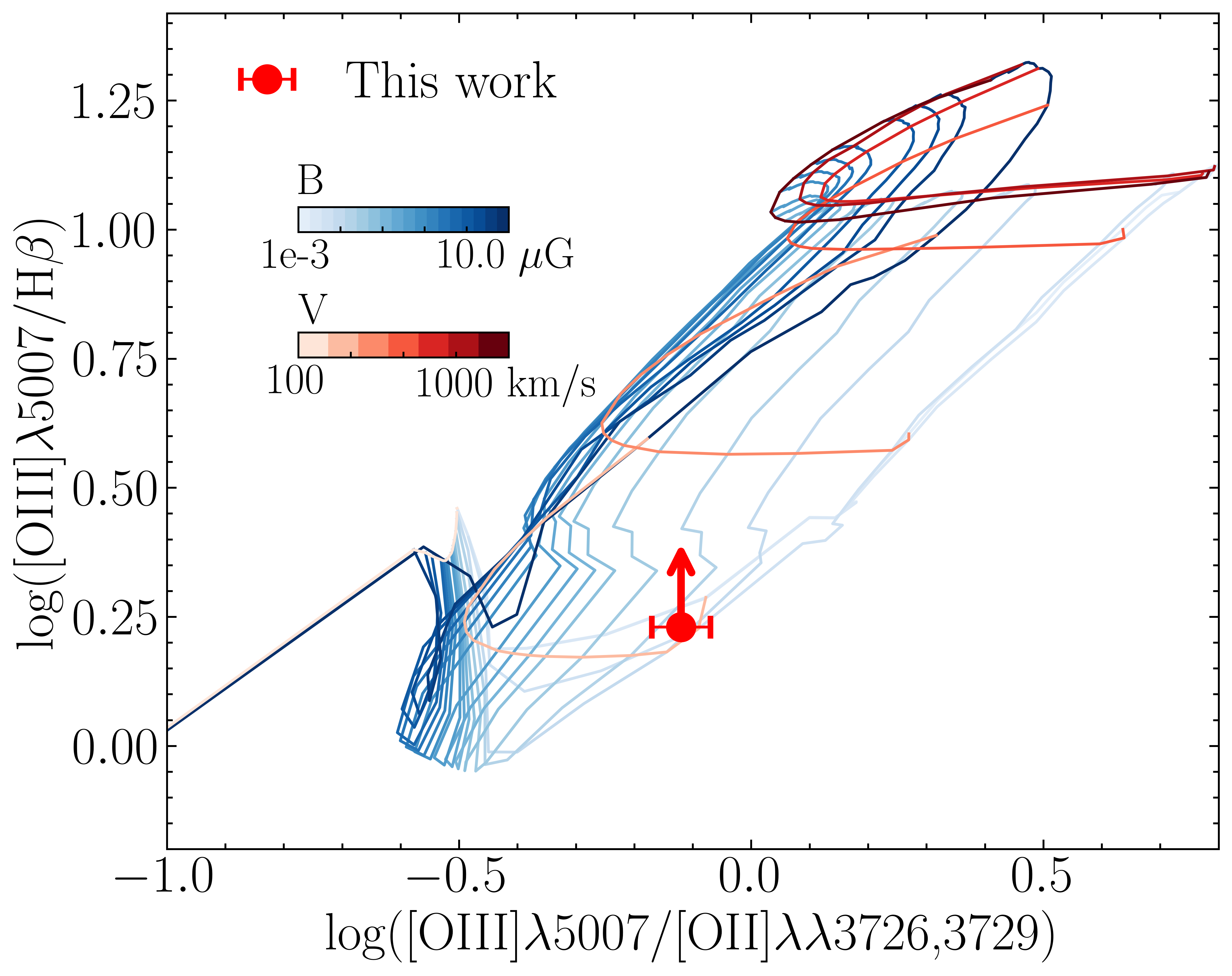} 
        \caption{
            Line ratio map of O32 and [\ion{O}{3}]/H$\beta$ from the MAPPINGS III library of fast radiative shock+precursor models with $n_{\rm{H}} = 0.1~\rm{cm^{-3}}$ and solar metallicity. The spectrum line ratio measured from LDSS3 with the red dot and the upper arrow, indicating the O32 ratio and the lower limit of [\ion{O}{3}]/H$\beta$. } 
    \label{shockgrids}
    \end{figure}

    So far, we have only considered the possibility of photoionization, where ionization by shocks generated by AGN, starburst, or merger activities could provide another solution for Oxyster. We investigate this possibility using the fast radiative shock libraries that cover a variety of gas density and metallicity based on the MAPPINGS III plasma modeling framework \citep{Allen-2008}. We have explored the pure shock model and the shock$+$precursor one that considers the pre-shock region affected by the upstream ionizing radiation escaped from the shock. Compared with the observed O32 line ratio and the lower limit of the [\ion{O}{3}]$\lambda5007/{\rm H}{\beta}$ ratio, we find that a shock$+$precursor model with solar metallicity, low gas density ($n = 0.1\,\rm{cm}^{-3}$ or $n = 0.01\,\rm{cm}^{-3}$), and shock speed $> 250$ km/s can provide a promising explanation (see Figure \ref{shockgrids} for the $n = 0.1\,\rm{cm}^{-3}$ model). In addition to the line ratios, assuming Oxyster is a $R=10$ kpc isotropic sphere of gas, such a shock$+$precursor model can also reproduce the observed [\ion{O}{2}] and [\ion{O}{3}] line luminosities based on the H$\beta$ flux density predicted by the shock model and the lower limit of the [\ion{O}{3}]/H$\beta$ line ratio. 

    To confirm or rule out shocks as the primary excitation mechanism, follow-up observations of the [\ion{S}{2}]$\lambda\lambda6717,6731$ doublet will be crucial for constraining the electron density. Additionally, observations of the [\ion{O}{3}]$\lambda4363$ auroral line, combined with [\ion{O}{3}]$\lambda5007$, will help determine the electron temperature, which can help separate the shock ionization scenario from the photoionization by AGN or starburst case, as shock can result in electron temperature higher than $T>10^{5.5}$ K (e.g., \citealt{Alarie-2019}).

    Meanwhile, the morphology of the ionized gas could be another consideration. Major merger events or the stellar winds from intense starbursts can create shocks that ionize the gas around these galaxies, resulting in an emission line `halo' accompanied by distinctive filaments \citep[e.g.,][]{Yoshida-2016, Rupke-2019, Nielsen-2020, Perrotta-2024}. While we currently do not have high-resolution emission line maps of Oxyster to investigate this, the fact that the emission line nebula is only on one side of the galaxy already hints at its difference with the above cases. Future narrow-band imaging or IFU observations with sufficient spatial resolution may further verify this. 

\subsection{The Single-arm Structure of the Host Galaxy and the Possible Merger Event}

    Another unique feature of the Oxyster system is that the host galaxy displays an unusual ``single-arm'' structure in the shorter wavelength images by {\it HST} (F606W and F814W) and {\it JWST} (F115W). In fact, the arm is almost the only visible feature in the rest-frame UV images, indicating a young starburst, which is confirmed by the SED fitting results. Interestingly, while the ``arm'' region has much lower dust attenuation than the galaxy core, it is not dust-free and shows an $A_V$ value similar to the average value of the whole galaxy, inconsistent with dust formed in the young starburst starting from $\sim 10^8$ yrs ago. This suggests the UV-bright ``arm'' reflects a recently induced starburst event superposed on the original stellar population. 
    
    \textbf{While our fits indicate significant dust attenuation ($A_{\rm V} \sim 2.8-3.7~\rm mag$), the IR luminosity remains modest based on the VLA catalog, since their IR-based SFR is only half of ours. This discrepancy may reflect a clumpy dust geometry or low dust covering fraction, allowing some UV photons to escape without being reprocessed \citep[e.g.,][]{2003ApJ...591.1049W, 2011ApJ...738..106W}. Alternatively, a cooler dust temperature or a different IR SED shape may reduce the inferred IR-based SFR \citep[e.g.,][]{2008MNRAS.388.1595D}.}

    Given the one-sided spiral arm structure, we speculate that the host galaxy has recently experienced a merger or interaction event that resulted in the infalling and compressing of gas in the arm region. An interaction could also help explain the existence of an extensive reservoir of cool gas around the host galaxy that feeds \emph{Oxyster}. The nearby galaxy (RA=10:00:38.278, DEC=+02:11:29.256) at the same redshift as the host galaxy, as reported in the COSMOS Spectroscopic Redshift Compilation \citep{Khostovan-2025}, supports this scenario. The proximity of the two galaxies suggests that the abundant CGM gas may have originated from tidal stripping during a recent interaction. Such interaction can also trigger a localized starburst. While this simple scenario may provide a coherent picture for Oxyster, it awaits further confirmation from future kinematic studies of the velocity field to trace potential merger-induced disturbances.
    
\subsection{Future Perspective}

    The serendipitous discovery of Oxyster at $z\sim 1$ uncovers an intriguing case for studying the Galactic ecosystem and galaxy-CGM connection right after the ``cosmic noon'' when star formation and AGN activities are declining. However, the narrow-band images and the current low-resolution spectra are insufficient to understand the physical processes fully. Future observations are essential to comprehensively understand the ionization state of the gas and the velocity field of the gas. Specifically, detecting prominent emission lines such as [\ion{N}{2}], [\ion{S}{2}], H$\beta$, and H$\alpha$ is necessary to constrain the ionization state of the gas using BPT diagrams. Furthermore, we anticipate that future observations will provide deeper insights into the gas velocity field and emission line distributions. IFUs that match the wavelength we require, such as VLT/MUSE and Keck/KCWI, are crucial for exploring these delicate structures.

    At the same time, the fact that we can discover Oxyster using narrow-band images within a moderate-sized extragalactic field around a star-forming galaxy suggests that a large population of similar objects may be waiting to be discovered. A systematic search with narrow- or medium-band filters over $>100\ {\rm deg}^2$ of the sky could result in a statistically meaningful sample of emission line nebula candidates around different types of galaxies, paving the way to assemble a valuable dataset to explore the cool CGM at the early Universe. Recently, studies based on JWST/NIRCam medium-band imaging have demonstrated its immense potential and effective capability to uncover the low-surface brightness CGM at high-$z$ \citep[e.g.,][]{Duncan2023MNRAS.522.4548D, Williams2023ApJS..268...64W, Suess2024ApJ...976..101S, Zhu2024arXiv240911464Z, Peng-2025, Solimano2025A&A...693A..70S}.
    However, the small field of view of JWST makes it difficult to build a large sample. Meanwhile, the arrival of Euclid images and spectroscopic redshifts from surveys like DESI will provide us with a large sample of $0.8<z<1.4$ star-forming galaxies with high-resolution images. If we believe interaction-induced starburst systems like Oxyster present an interesting opportunity to directly image the cool gas in the CGM using extended emission line nebulae, we could identify a sample of similar cases for narrow-band or IFU follow-up observations.

\section{Conclusion}
    \label{sec:conclude}

    This paper presents the discovery of a giant emission line nebula, `Oxyster', at z=0.924, extending to 30 kpc from a star-forming host galaxy. The spatially resolved [\ion{O}{2}] and [\ion{O}{3}] emissions were first identified using the public narrow-band images of Subaru HSC CHORUS Survey and then confirmed with follow-up spectroscopic observations using Magellan-LDSS3. 
    
    Oxyster is characterized by a high [\ion{O}{2}] luminosity ($9.24\pm0.09 \times 10^{41}\ {\rm erg/s}$) and a lower O32 ratio than 1.0, suggesting a relatively low ionization state, while the lower limit of [\ion{O}{3}]5007/H$\beta$ suggests a hard ionization source. At fixed [\ion{O}{3}] luminosity, Oxyster sits above the luminosity-size relation for the extended narrow-line regions of low-redshift AGNs but is very similar to the famous `Hanny's Voorwerp'. 

    Through SED fitting, we find that the host galaxy is a starburst galaxy with $\sim 2-5\times 10^{10}\ {\rm M_\odot}$ stellar mass and $\sim 100-140\ {\rm M_\odot/yr}$ star formation rate, depending on the SFH model. The host galaxy exhibits a one-sided spiral arm structure, brighter than the core in the rest-frame UV to optical-blue bands. The arm is characterized by a young stellar population ($\sim 2-4\times 10^{9}\ {\rm M_\odot}$) formed in a recent starburst with $\sim 50 \ {\rm M_\odot/yr}$ star formation rate, which could be triggered by a merger event. Regardless of the SFH model, the galaxy and the arm are above the star-forming main sequence at $z=1$. The galaxy does not show strong evidence of AGN activity, with an upper limit of 5\% AGN contribution from the SED fitting. The VLA 3GHz catalog classifies the host galaxy as a mid-to-low luminosity AGN based on the radio excess relative to the expected value supported by an IR-based SFR value. The X-ray luminosity upper limit is $\rm{log}~L_{\rm{X}} < 41.9~\rm{erg^{-1}~s^{-1}}$.
    
    Oxyster's ionization mechanism is still under debate, with the possibility of a combination of starburst and AGN contribution. We modeled the nebula as a cloud 10 kpc from the radiation field with 10 kpc thickness using \texttt{CLOUDY}. The result is quantitatively consistent with the observed line ratios, suggesting a low-luminosity AGN with a bolometric luminosity of $1\times10^{42} \sim 1\times10^{43}$ erg/s. The X-ray luminosity is comparable to the upper limit derived from the Chandra detection limit. The nebula can also be ionized by starburst radiation alone, with a star formation rate of $\sim 97 \ {\rm M_\odot/yr}$, close to the SED fitting results. In addition, shock excitation models reproduce the observed line ratios and luminosities but require high shock velocities. The ionization mechanism can be further constrained by future observations of additional emission lines and the velocity field of the gas.
    
    While being a single object, Oxyster highlights the potential of systematically studying CGM in the early Universe and the importance of understanding the galaxy ecosystem at that time. The unique features of Oxyster, such as the low O32 ratio, the ``missing'' Balmer emission lines, and the asymmetric morphology of the host galaxy, all beg for further discussion. While the majority of the attention around CGM has been focused on high-redshift AGNs, Oxyster demonstrates the importance of studying the CGM around more ``normal'' star-forming galaxies, which can be achieved through systematic narrow/medium-band imaging surveys.

    
    \textbf{In the interest of transparency and reproducibility, we provide the plotting scripts used to generate the figures in this paper in a public GitHub repository: \url{https://github.com/SunnySco/giant-nebula}.}
    
\section*{Acknowledgments}

    The authors thank Yunjing Wu, Ben Wang, Xiaojing Lin, and Daming Yang for their helpful instructions on data analysis. 
    We also thank Zhiyu Zhang, Sebastiano Cantalupo, Lu Shen, Yanmei Chen, Haibin Zhang, Rhythm Shimakawa, Yifei Jin, and Ruqiu Lin for insightful discussions and Zechang Sun for maintaining the \emph{dodo} Server at Tsinghua University. 
    PJL appreciates the support from other lovely members of the Guangtou Group.

    SH acknowledges support from the National Natural Science Foundation of China Grant No. 12273015, No. 12433003, and the China Crewed Space Program through its Space Application System. 
    ZC and ML acknowledge support from the National Key R\&D Program of China (grant no. 2023YFA1605600) and Tsinghua University Initiative Scientific Research Program (No. 20223080023).

    \textbf{This research is based on observations made with the NASA/ESA Hubble Space Telescope and NASA/ESA/CSA James Webb Space Telescope. HST data were obtained from the Space Telescope Science Institute, which is operated by the Association of Universities for Research in Astronomy, Inc., under NASA contract NAS 5–26555. JWST data were obtained from the Mikulski Archive for Space Telescopes at the Space Telescope Science Institute, which is operated by the Association of Universities for Research in Astronomy, Inc., under NASA contract NAS 5-03127 for JWST. These observations are associated with program \#1727. The authors acknowledge the COSMOS-Web team led by coPIs (J. Kartaltepe and C. Casey) for developing their observing program with a zero-exclusive-access period.}
    
    Some of the data products presented herein were retrieved from the Dawn JWST Archive (DJA). DJA is an initiative of the Cosmic Dawn Center (DAWN), which is funded by the Danish National Research Foundation under grant DNRF140. 
    
    \textbf{\facilities{HST(ACS,WFC3), CFHT(Mega-Prime, WIRCam), Subaru(HSC), JWST(NIRCam)}}
    
    \software{    \texttt{Astropy} \citep{Astropy-1, Astropy-2, Astropy-3}, 
    \texttt{PypeIt} \citep{Prochaska-2020a},
    \texttt{PSFEx} \citep{Bertin-2011},
    \texttt{SExtractor} \citep{Bertin-1996},
    \texttt{WebbPSF} \citep{Perrin-2012},
    \texttt{pypher} \citep{Boucaud-2016},
    \texttt{photutils} \citep{photoutils},
    \texttt{mwdust} \citep{Bovy-2016},
    \texttt{extinction} \citep{barbary-2017},
    \texttt{Prospector} \citep{Leja-2019, Johnson-2021},
    \texttt{Bagpipes} \citep{Carnall-2018},
    \texttt{CLOUDY} \citep{Chatzikos-2023},
    \texttt{MyFilter} \citep{MyFilter}.
    }

\vspace{5mm}
\appendix
\section{Multi-band images of \emph{Oxyster}}
\label{sec:appendix}
In this appendix, we show the $10\times10~\rm{arcsec}^2$ cutouts of multiwavelength images, which we use to perform the SED fitting, listed in Table \ref{photodata}.

\begin{figure}[htbp]
\centering
\includegraphics[width=0.9\textwidth]{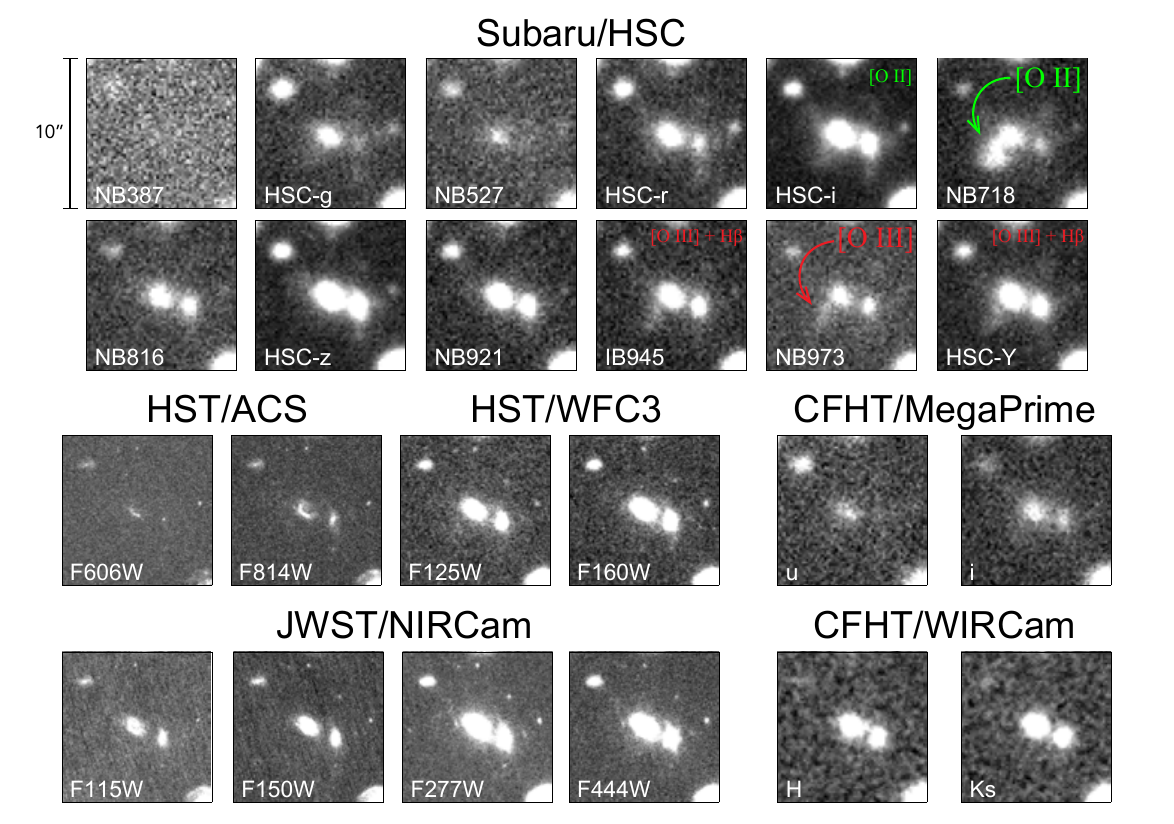}
\caption{Image cutouts within $10\times10~\rm{arcsec}^2$ taken by Subaru, HST, JWST, and CFHT listed in Table \ref{photodata}.} 
\label{bands-cutout}
\end{figure}

\bibliography{references}{}
\bibliographystyle{aasjournal}

\end{document}